\documentclass[aps,prb,preprint,onecolumn,citeautoscript,superscriptaddress,footinbib,
eqsecnum]{revtex4-1}  
\usepackage{amsmath,amssymb,bm,bbm} 
\usepackage{graphicx}  
\usepackage{color} 
\usepackage[papersize={8.5in,11in}]{geometry}
\usepackage[colorlinks=true]{hyperref}  
\hypersetup{
    bookmarks=true,         
    unicode=false,          
    pdftoolbar=true,        
    pdfmenubar=true,        
    pdffitwindow=false,     
    pdfstartview={FitH},    
    pdftitle={Insulators and metals with topological order and discrete symmetry breaking},    
    pdfauthor={Subir Sachdev and Shubhayu Chatterjee},     
    pdfsubject={},   
    pdfcreator={},   
    pdfproducer={}, 
    pdfkeywords={} {} {}, 
    pdfnewwindow=true,      
    colorlinks=true,       
    linkcolor=magenta, 
    citecolor=blue,        
    filecolor=magenta,      
    urlcolor=blue           
} 

\usepackage{amsfonts}
\usepackage{upgreek}
\usepackage{slashed}
\usepackage{latexsym}
\usepackage{todonotes}

\newcommand{\beq}{\begin{equation}}
\newcommand{\eeq}{\end{equation}}
\def\bea{\begin{eqnarray}}
\def\eea{\end{eqnarray}}
\def \k{{\bm k}}
\def \Q{{\bm Q}}
\def \vr{{\bm v}_\rho}
\def \on{{\omega_n}}
\def \sa{{\sigma^a}}
\DeclareMathOperator{\Tr}{Tr}
\newcommand{\nn}{\nonumber \\}

\begin{document}

\title{Insulators and metals with topological order\\ and discrete symmetry breaking}

\author{Shubhayu Chatterjee}
\affiliation{Department of Physics, Harvard University, Cambridge MA 02138, USA}

\author{Subir Sachdev}
\affiliation{Department of Physics, Harvard University, Cambridge MA 02138, USA}
\affiliation{Perimeter Institute for Theoretical Physics, Waterloo, Ontario, Canada N2L 2Y5}

\date{February 28, 2017 
\\
\vspace{0.4in}}

\begin{abstract}
Numerous experiments have reported discrete symmetry breaking in the high temperature pseudogap phase of the hole-doped cuprates,
including breaking of one or more of lattice rotation, inversion, or time-reversal symmetries. In the absence of translational symmetry breaking or
topological order, these
conventional order parameters cannot explain the gap in the charged fermion excitation spectrum in the anti-nodal region.
Zhao {\em et al.\/} (Nature Physics {\bf 12}, 32 (2016)) and Jeong {\em et al.\/} (Nature Communications {\bf 8}, 15119 (2017))
have also reported inversion and time-reversal symmetry breaking in insulating 
Sr$_2$IrO$_4$ similar to that in the metallic cuprates, but co-existing with N\'eel order. 
We extend an earlier theory of topological order in insulators and metals, in which 
the  topological order combines naturally with the breaking of these conventional
discrete symmetries. We find translationally-invariant states with topological order co-existing with both
Ising-nematic order and spontaneous charge currents. The link between the discrete broken symmetries and the topological-order-induced pseudogap explains why the broken symmetries do not survive in the confining phases without a pseudogap at large doping.
Our theory also connects to the O(3) non-linear sigma model and $\mathbb{CP}^1$ descriptions of quantum fluctuations of the N\'eel order.
In this framework, the optimal doping criticality of the cuprates is primarily associated with the loss of topological order.
\end{abstract}
\maketitle

\section{Introduction}
\label{sec:intro}

Experimental studies of the enigmatic high temperature `pseudogap' regime of the hole-doped cuprate compounds 
have reported numerous possible discrete symmetry breaking order parameters \cite{2002PhRvL..88m7005A,2006PhRvL..96s7001F,Hinkov597,2008arXiv0805.2959L,2008PhRvL.100l7002X,2010Natur.463..519D,2010Natur.468..283L,2010Natur.466..347L,2014PhRvL.112n7001L,2015NatCo...6E7705M,2016NatPh..12...32Z,2016arXiv161108603Z,2017arXiv170106485J}.
There is evidence for lattice rotation symmetry breaking, interpreted in terms of an Ising-nematic order \cite{kfe98}, and for one or both
of inversion and time-reversal symmetry breaking, usually interpreted in terms of Varma's current loop order \cite{1997PhRvB..5514554V,2002PhRvL..89x7003S,2003PhRvB..67e4511S}. Both of these orders have the full translational symmetry of the square lattice, and cannot, by themselves, be responsible for gap in the charged fermionic spectrum near the `anti-nodal' points ($(\pi, 0)$ and $(0,\pi)$) of the
square lattice Brillouin zone.

An interesting and significant recent development has been the observation of inversion \cite{2016NatPh..12...32Z} and time-reversal \cite{2017arXiv170106485J} symmetry breaking in the iridate compound
Sr$_2$Ir$_{1-x}$Rh$_x$O$_4$; 
Ref.~\onlinecite{2011PhRvL.106m6402W} has shown that this iridate is
described by a one-band Hubbard model very similar to that for the cuprates. The inversion symmetry breaking  is strongest in the insulator at $x=0$ where it co-exists with N\'eel order; at non-zero $x$, both orders persist, but the discrete order is present at higher temperatures. Motivated by the similarities in the light and neutron scattering signatures 
between the cuprate and iridate compounds, we will present here 
a common explanation based upon the quantum fluctuations of antiferromagnetism.

Long-range N\'eel order (which breaks translational symmetry) can 
clearly be the origin of a gap in the charged fermionic spectrum at the anti-nodes. In the traditional
spin density wave theory of the quantum fluctuations of the N\'eel order \cite{hertz}, there is a transition to a state without
N\'eel order, with full translational symmetry, a large Fermi surface, and no anti-nodal gap.
However, the anti-nodal gap can 
persist into the non-N\'eel phase \cite{CScomment}
when the resulting phase has topological order \cite{FFL,TSMVSS04,APAV04,2015PNAS..112.9552P,2012PhRvB..85s5123P} (see footnote\footnote{
Topological order is defined by the presence of ground state degeneracy of a system on a torus. More precisely, on a torus of size $L$, the lowest energy states have an energy difference which is of
order $\exp (-\alpha L)$ for some constant $\alpha$. Topological order can also be present in gapless states, including those with Fermi surfaces \cite{FFL,TSMVSS04}. In such states, the non-topological gapless excitations
have an energy of order $1/L^z$ (for some positive $z$) above the ground state on the torus,
and so can be distinguished from the topologically degenerate states. Topological order is required
for metals to have a Fermi surface volume distinct from the Luttinger volume \cite{TSMVSS04}, and hence to have a
`pseudogap'.} for the precise definition of topological order, and the review in Ref.~\onlinecite{DCSS16}). We shall use topological order as the
underlying mechanism for the pseudogap. 
Moreover, early studies of spin liquid insulators with $\mathbb{Z}_2$ topological order showed that there can be a non-trivial interplay between
topological order and the breaking of conventional discrete symmetries. 
The $\mathbb{Z}_2$ spin liquid obtained 
in Refs.~\onlinecite{NRSS91,SSNR91} co-existed with Ising-nematic order: this
was a consequence of the $p$-wave pairing of bosonic spinons.
A similar interplay
with time-reversal and inversion symmetries was discussed by Barkeshli {\it et al.}~\cite{2013PhRvB..87n0402B}, using higher angular momentum pairing
of fermionic spinons.
Here we shall use the formalism of Refs.~\onlinecite{SS09,DCSS15b,DCSS16}
to generalize the state \cite{NRSS91,SSNR91} with $\mathbb{Z}_2$ 
topological order to also allow for the breaking of inversion and time-reversal symmetries, both
in the insulator and the metal. We will find states with spontaneous charge currents
(see Fig.~\ref{fig:currents2}) and topological order, one of which (Fig.~\ref{fig:currents2}a) also has the Ising-nematic order observed in 
experiments \cite{2002PhRvL..88m7005A,Hinkov597,2010Natur.463..519D,2010Natur.466..347L}. 

The association between topological order and discrete broken symmetries implies that the broken symmetries will not be present in the confining phases at larger doping. This is an important advantage of our approach over more conventional excitonic condensation theories of broken symmetries. In the latter approaches there is no strong reason to connect the disappearance of the pseudogap with vanishing of the symmetry order parameter. 

We will begin in Section~\ref{sec:o3} by a semi-classical treatment of the quantum fluctuations of antiferromagnetism \cite{CHN1,CHN2,Haldane88}
using the O(3) non-linear sigma model. 
In the insulator, this approach has been successfully used to describe the thermal fluctuations of the N\'eel order, and also the adjacent
quantum phase without N\'eel order; the latter was argued to have valence bond solid (VBS) order \cite{NRSS89,NRSS90}, 
and is accessed across a deconfined
quantum critical point \cite{senthil1,senthil2}. Here, we will identify an
order parameter, ${\bm O}$, for inversion and time-reversal symmetry
breaking in terms of the fields of the O(3) sigma model.

It is also useful to formulate the semi-classical treatment using the $\mathbb{CP}^1$ model
for bosonic, fractionalized spinons coupled to a U(1) gauge field. In these terms, an order parameter ${\bm O}$ for inversion and time-reversal symmetry
breaking turns out to be the cross product of the emergent U(1) electric and magnetic fields. The $\mathbb{CP}^1$ formulation yields
an effective gauge theory, in Eq.~(\ref{Sa}) for quantum phases with spontaneous charge currents, 
but without N\'eel order.

\begin{figure}
\begin{center}
\includegraphics[height=6cm]{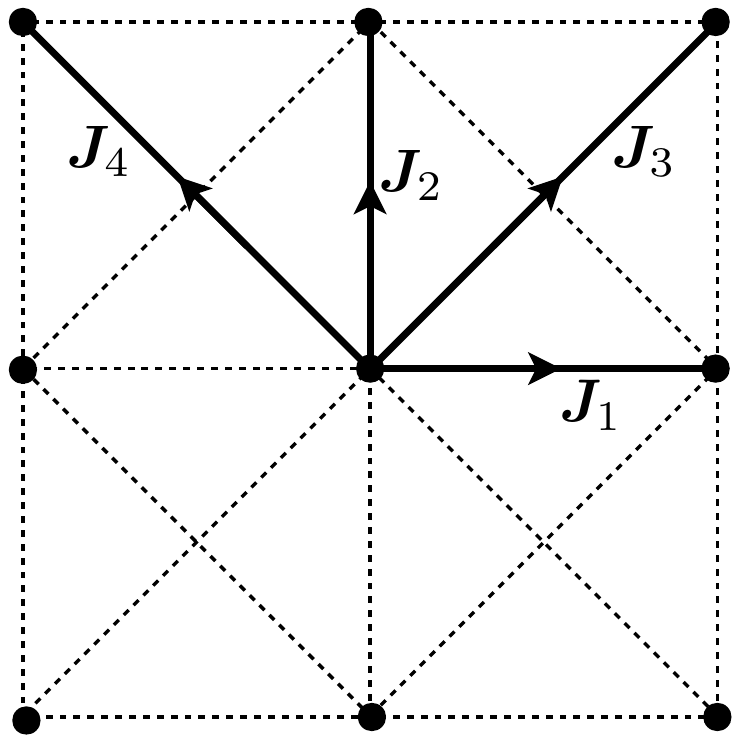}
\end{center}
\caption{Definitions of currents on the square lattice with first and second neighbor
hopping. The filled circles are the sites of the Cu atoms in the cuprates. 
Shown above are the 4 currents
${\bm J}_\rho$ from the central lattice site. These currents obey Eq.~(\ref{bloch3}) when the translational
symmetry of the square lattice is preserved.}
\label{fig:currents1}
\end{figure}

\begin{figure}
\begin{center}
\includegraphics[height=6.5cm]{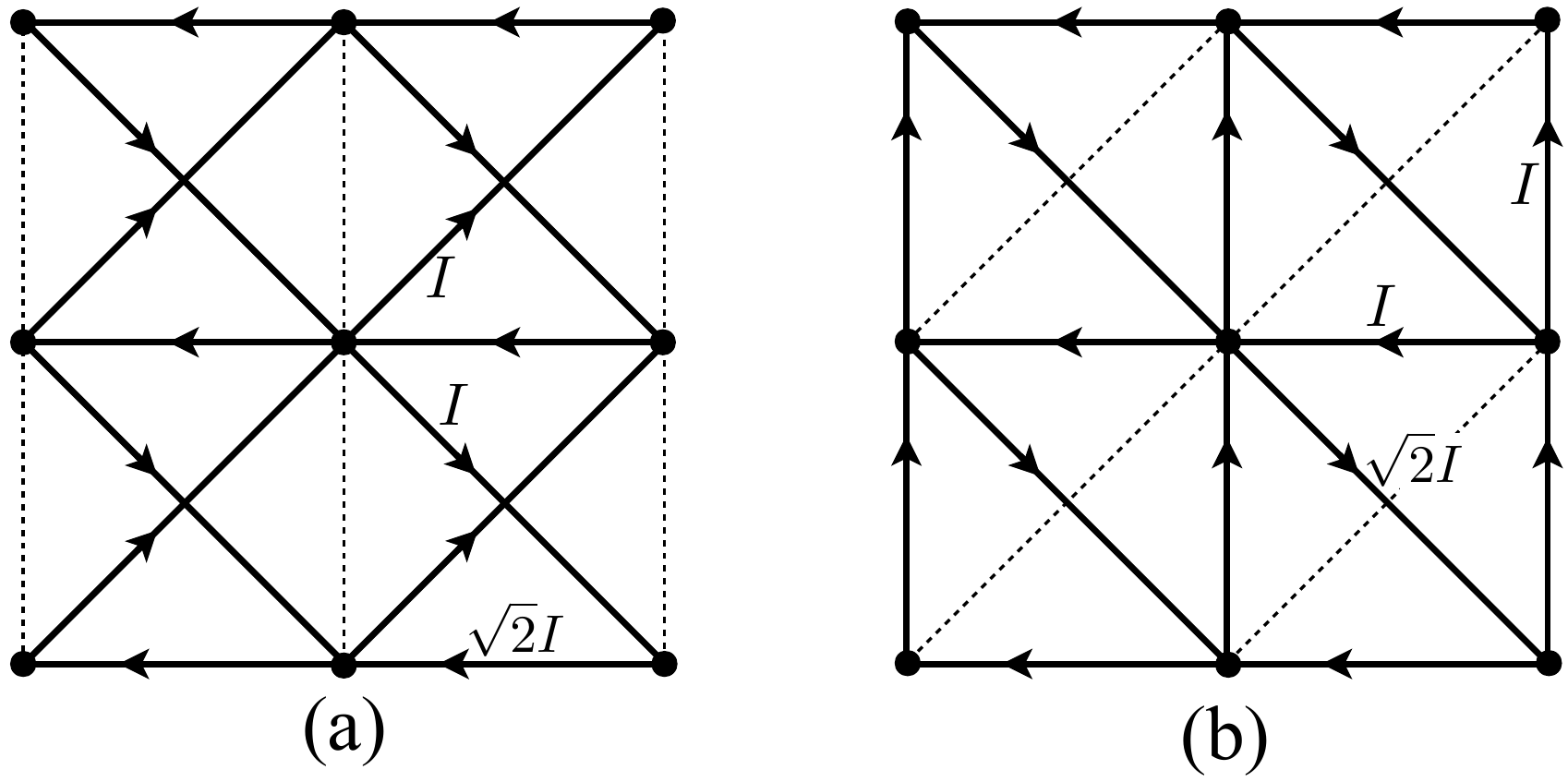}
\end{center}
\caption{Currents on the links in two classes of states with broken time reversal and inversion symmetries.
The states have the full translational symmetry of the
square lattice, and the magnitude of the current is noted on some links. The current vanishes on the dashed lines. In (a), the currents have the magnitudes $|{\bm J}_1| = \sqrt{2} I$, $|{\bm J}_2| = 0$, $|{\bm J}_3| = I$, $|{\bm J}_4| = I$, and the order parameter ${\bm O} \sim (-1,0)$.
In (b), the currents have the magnitudes $|{\bm J}_1| =  I$, $|{\bm J}_2| = I$, $|{\bm J}_3| = 0$, $|{\bm J}_4| = \sqrt{2} I$, and the order parameter ${\bm O} \sim (1,-1)$. The state in (a) has Ising-nematic order $\mathcal{N}_1$ non-zero, while the state in (b) has Ising-nematic order
$\mathcal{N}_2$ non-zero (see Eq.~(\ref{defnematic})). Experiments on the cuprates \cite{2002PhRvL..88m7005A,Hinkov597,2010Natur.463..519D,2010Natur.466..347L} observe the Ising-nematic order $\mathcal{N}_1$.
}
\label{fig:currents2}
\end{figure}

Formally, a model expressed in terms of spins alone has no charge fluctuations, and so has vanishing
electromagnetic charge current, ${\bm J}=0$. 
However, in practice, every spin model arises from an underlying Hubbard-like model, in which
states suppressed by the on-site repulsion $U$ are eliminated by a canonical transformation. If we undo this canonical transformation, we can
expect that a suitable multi-spin operator will induce a non-zero ${\bm J}$ at some order in the $1/U$ expansion.
As ${\bm O}$ has the same symmetry signature as ${\bm J}$, we can expect that 
a state with $\langle {\bm O} \rangle$ non-zero will also have $\langle {\bm J} ({\bm r}) \rangle$ non-zero.
We will examine states in which $\langle {\bm O} \rangle$ is independent of ${\bm r}$ in the continuum limit, 
so that translational symmetry is preserved. However, by Bloch's
theorem \cite{Bohm49,1996JPSJ...65.3254O}, we must have
\beq
\int d^2 r \langle {\bm J} ({\bm r}) \rangle = 0, \label{bloch}
\eeq
and so ${\bm J}$ cannot be ${\bm r}$
independent. If we want to preserve translational symmetry, the resolution is that there
will be intra-unit cell variations in $\langle {\bm J} ({\bm r}) \rangle$ to retain compatibility with Bloch's theorem. In a tight-binding model with one site per unit cell, we 
label each unit cell by a site label, $i$, and a link label $\rho$ so that the combination
$(i, \rho)$ identifies the complete set of lattice links, with no double counting. So from each lattice site $i$, there
are set of vectors ${\bm v}_\rho$ connecting $i$ to its neighboring sites, and both ${\bm v}_\rho$ and
$-{\bm v}_\rho$ are not members of this set: see Fig.~\ref{fig:currents1}.
In this setup, Bloch's theorem states that
\begin{equation}
\sum_{\rho} \langle {\bm J}_\rho \rangle  = 0. \label{bloch3}
\end{equation}
where ${\bm J}_\rho$ is the current along the ${\bm v}_\rho$ direction. 
Note that Eq.~(\ref{bloch3}) is a stronger statement than current conservation because the sum over
$\rho$ does not include all links connected to site $i$, only half of them.
Eq.~(\ref{bloch3}) is equivalent to the statement that there are current `loops', and these
are clearly possible even in a single-band model \cite{2004PhRvB..69x5104S,2008PhRvL.100b7003B}. 
In the presence of a ${\bm r}$-independent ${\bm O}$ condensate, we can write by symmetry that (to linear
order in the broken symmetry)
\beq
\left\langle J_{\rho p} \right\rangle  = K^\rho_{pp'} \left \langle O_{p'} \right\rangle, \label{JKO}
\eeq
where $p,p'=x,y$ are spatial indices, and $K_{pp'}^\rho$ is a response function 
obtained in the $1/U$ expansion which respects
all square lattice symmetries. Compatibility with Bloch's theorem requires that
\beq
\sum_\rho K_{ij}^\rho  = 0, \label{bloch2}
\eeq
and there are no conditions on the value of $\langle {\bm O} \rangle$.

We will turn to an explicit treatment of the charged excitations, and a computation of ${\bm J}_{\rho}$ in Section~\ref{sec:lattice}: our results there do obey Eqs.~(\ref{bloch}) and (\ref{bloch2}). 
Section~\ref{sec:lattice} 
will present a lattice formulation in which the U(1) gauge field of the $\mathbb{CP}^1$ model is embedded in a SU(2) lattice gauge theory \cite{SS09,DCSS15b,DCSS16,2016PhRvB..94k5147S}. This lattice gauge theory 
has the advantage of including all Berry phases and charged fermionic excitations,  and also for allowing an eventual
transition into a conventional Fermi liquid state at high enough doping. Our interest here will be in insulating and metallic states at lower
doping, which have topological order and a gap to charged fermionic excitations in the anti-nodal region. At the same time we shall show that, with an appropriate
effective action, there can be a background modulated gauge flux under which {\it gauge-invariant} 
observables remain translationally invariant
but break one or more of inversion, time-reversal and lattice rotation symmetries. Our computations will demonstrate the presence of spontaneous 
charge currents obeying Eq.~(\ref{bloch3}) in states which break both inversion and time-reversal, but preserve translation. The two classes of spontaneous current patterns we find are shown in Fig.~\ref{fig:currents2}.
Note that the product of time-reversal and inversion is preserved in these states. The state in Fig.~\ref{fig:currents2}a has $\langle {\bm O} \rangle \sim (-1,0)$, while the state in Fig.~\ref{fig:currents2}b has $\langle {\bm O} \rangle \sim (1,-1)$. Both states belong to separate quartets
of equivalent states (with $\langle {\bm O} \rangle \sim (\pm 1,0), (0, \pm 1)$ and 
$\langle {\bm O} \rangle \sim (\pm 1,\pm 1)$) which can be obtained from them by symmetry operations.
Both states also break an Ising-nematic symmetry. In general, on the square lattice, we can define
two Ising-nematic order parameters, which are invariant under both inversion and time-reversal, but
not under lattice rotation symmetries. In terms of ${\bm O}$, these order parameters are
\begin{equation}
    \mathcal{N}_1 = O_x^2 - O_y^2 \quad , \quad \mathcal{N}_2 = O_x O_y . \label{defnematic}
\end{equation}
The state in Fig.~\ref{fig:currents2}a has only $\langle \mathcal{N}_1 \rangle \neq 0$, 
while the state in Fig.~\ref{fig:currents2}b has only $\langle \mathcal{N}_2 \rangle \neq 0$.
We note that the state in Fig.~1b of Simon and Varma \cite{2002PhRvL..89x7003S} in a two band model has
the same symmetry as the state in our Fig.~\ref{fig:currents2}b, and also the spontaneous current 
states considered in Refs.~\onlinecite{2004PhRvB..69x5104S,2008PhRvL.100b7003B}. 
The state in our Fig.~\ref{fig:currents2}a appears
to not have been considered earlier: it has the same Ising-nematic order observed in experiments in the 
cuprates \cite{2002PhRvL..88m7005A,Hinkov597,2010Natur.463..519D,2010Natur.466..347L}.

Finally, we note that our results are also easily extended to states with long-range antiferromagnetic
order by condensing the spectator bosonic spinons.

\section{O(3) non-linear sigma and $\mathbb{CP}^1$ models}
\label{sec:o3}

The familiar O(3) model describes quantum fluctuations of the unit vector $\vec{n} ( {\bm r}, \tau)$, representing the local
antiferromagnetic order, with action
over space, ${\bm r}$, and imaginary time $\tau$
\beq
\mathcal{S}_{\vec{n}} = \frac{1}{2 g} \int d^2 r d \tau \left( \partial_\mu \vec{n} \right)^2 ,
\eeq
where $\mu$ extends over the 3 spacetime indices, and $g$ is a coupling constant. For our purposes, we need the symmetry transformation
properties of the operator $\vec{n}$ and its canonically conjugate angular momentum $\vec{L}$; the latter is also interpreted as the 
conserved ferromagnetic moment \cite{CHN2}. We list these transformations properties in Table~\ref{table:o3}.

Table~\ref{table:o3} also shows the symmetry transformations of charge current ${\bm J}$. 
Formally, a model expressed in terms of spins alone has no charge fluctuations, and so we will have ${\bm J}=0$. 
\begin{table}
\begin{tabular}{|c|c|c|c|c|c|}
\hline
~~ & ~~$\mathcal{T}$~~ & ~~$T_x$~~ & ~~$T_y$~~ & ~~$I_x$~~ & ~~$I_y$~~ \\
\hline
~~$\vec{n}$~~ & $-$ & $-$ & $-$ & $+$ & $+$ \\
$\vec{L}$ & $-$ & $+$ & $+$ & $+$ & $+$\\
$ e_x $ & $+$ & $-$ & $-$ & $-$ & $+$ \\ 
$ e_y $ & $+$ & $-$ & $-$ & $+$ & $-$ \\ 
$b$ & $-$ & $-$ & $-$ & $-$ & $-$ \\
$J_x$ & $-$ & $+$ & $+$ & $-$ & $+$ \\
$J_y$ & $-$ & $+$ & $+$ & $+$ & $-$ \\
\hline
\end{tabular}~\\~\\
\caption{Symmetry signatures of various fields under time reversal ($\mathcal{T}$), translation by a square lattice spacing along the $x$
($T_x$) and $y$ ($T_y$) directions, and reflections about a square lattice site involving $x \rightarrow -x$ ($I_x$) or $y \rightarrow -y$ ($I_y$).}
\label{table:o3}
\end{table}
However, in practice, every spin model arises from an underlying Hubbard-like model, in which
states suppressed by the on-site repulsion $U$ are eliminated by a canonical transformation. If we undo this canonical transformation, we can
expect that a suitable multi-spin operator will couple linearly to ${\bm J}$ at some order in the $1/U$ expansion; naturally, we need this
multi-spin operator to have the same symmetry signature as ${\bm J}$.
We therefore use Table~\ref{table:o3} to find the simplest such combination of $\vec{n}$ and $\vec{L}$;
the needed operator turns out to be
\beq
{\bm O} = \vec{L} \cdot ( \vec{n} \times {\bm \nabla} \vec{n} ). \label{Oo3}
\eeq
We will therefore be interested in states in which $\langle {\bm O} \rangle$ is non-zero and independent of ${\bm r}$. 
In magnetically ordered states, $\langle {\bm O} \rangle$ is non-zero for a `canted spiral' in which the spins precess around the base of a cone along a fixed spatial direction \cite{PhysRevLett.108.087205}: a non-zero 
$\vec{n} \times {\bm \nabla} \vec{n}$ corresponds to a spin spiral, which must cant to introduce a non-zero $\vec{L}$.
However, our interest here is in states in which $\langle {\bm O} \rangle$ is non-zero without long-range magnetic order,
in which 
$\langle \vec{n} \rangle =0$ and $\langle \vec{L} \rangle = 0$. For example, we can add to $S_{\vec{n}}$ an
effective potential $V ({\bm O})$ which is invariant under all symmetries, and a suitable $V ({\bm O})$ will induce an ${\bm O}$ condensate.
In Section~\ref{sec:lattice}, we will present specific lattice models for which such condensates arise.
In any such state with an ${\bm O}$ condensate, we can also expect that $\langle {\bm J} ({\bm r}) \rangle$ is also non-zero, and obeys Eqs.~(\ref{JKO}) and (\ref{bloch2}).

Let us now turn to the $\mathbb{CP}^1$ model. This is expressed in terms of bosonic spinons, $z_\alpha$, with $\alpha = \uparrow, \downarrow$
and $|z_\uparrow|^2 + |z_\downarrow|^2 = 1$, related to the antiferromagnetic order by
\beq
\vec{n} = z_\alpha^\ast \vec{\sigma}_{\alpha\beta} z_\beta^{\vphantom \dagger},
\eeq
where $\vec{\sigma}$ are the Pauli matrices. The action for the $\mathbb{CP}^1$ model has an emergent U(1) gauge field $a_\mu = (a_\tau, {\bm a})$:
\beq
\mathcal{S}_z = \frac{1}{g} \int d^2 r d \tau\, |(\partial_\mu - i a_\mu) z_\alpha|^2 . \label{Sz}
\eeq
We define the associated emergent electric and magnetic fields by, as usual by
\beq
{\bm e} = \partial_\tau {\bm a} - {\bm \nabla} a_\tau \quad , \quad b =  \hat{\bm z} \cdot ( {\bm \nabla} \times {\bm a} ) ,
\eeq
where $\hat{\bm z}$ is a unit vector orthogonal to the square lattice in the $x$-$y$ plane. 
These gauge-invariant fields are connected to the topological charge of the O(3) order parameter $\vec{n}$ via
\beq
{\bm e} = \frac{1}{2} \vec{n} \cdot \left( \partial_\tau \vec{n} \times {\bm \nabla} \vec{n} \right) \quad , \quad 
b = \frac{1}{2} \vec{n} \cdot \left( \partial_x \vec{n} \times \partial_y \vec{n} \right). \label{ebn}
\eeq
We can now use (\ref{ebn}) to deduce the symmetry signatures of ${\bm e}$ and $b$, and the results were shown in Table~\ref{table:o3}.
Finally, as in the O(3) formulation, we now search for a combination of ${\bm e}$ and $b$ which has the same symmetry signature as
the charge current; the simplest possibility is
\beq
{\bm O} = {\bm e} \times (b \, \hat{\bm z}) . \label{Ocp1}
\eeq
Note that the operators in Eqs.~(\ref{Oo3})
and (\ref{Ocp1}) are not equal to each other: they are distinct representations with the same symmetry signature. The connection between ${\bm O}$
and the charge current ${\bm J}$ in Eq.~(\ref{JKO}) also applies to Eq.~(\ref{Ocp1}). Also at this order, ${\bm O}$
is equal to the conserved Poynting vector of the gauge field, but we do not expect possible higher order terms in 
${\bm O}$ to yield a conserved quantity. Also, as for the O(3) model, we can add a suitable potential $V({\bm O})$
to the $\mathbb{CP}^1$ action $\mathcal{S}_z$ in Eq.~(\ref{Sz}) and induce a phase with an ${\bm O}$ condensate.

The advantage of the $\mathbb{CP}^1$ formulation is that we can now write down an effective action for the phase without N\'eel order,
where the $z_\alpha$ spinons are gapped. We integrate out the $z_\alpha$ spinons and generate an effective action for the U(1) gauge field $a_\mu$
in the state where ${\bm O}$ is condensed; using gauge invariance and symmetries, the imaginary time action has the form
\beq
\mathcal{S}_a = \int d^2 r d \tau \left[ \frac{\gamma_1}{2} (\partial_\tau a_i - \partial_i a_\tau)^2 + \frac{\gamma_2}{2} (\partial_x a_y - \partial_y a_x)^2
+ i \Gamma_i (\partial_\tau a_i - \partial_i a_\tau)(\partial_x a_y - \partial_y a_x)\right], \label{Sa}
\eeq
where $\gamma_{1,2}$ are coupling constants. The novel feature is the last term which has a co-efficient proportional to $\langle {\bm O} \rangle$
\beq
{\bm \Gamma} \propto \hat{\bm z} \times \langle {\bm O} \rangle;
\eeq
this term leads to a relatively innocuous modification of the gauge field propagator from the familiar relativistic form.
By itself, the U(1) gauge theory $\mathcal{S}_a$ is unstable to confinement by 
the proliferation of monopoles and the appearance of VBS order \cite{NRSS89}.
However, topological order can be stabilized if there are Fermi surfaces of U(1) charged fermions \cite{2004PhRvB..70u4437H,2008NatPh...4...28K}
which suppress monopoles. Alternatively, $\mathbb{Z}_2$ topological order can be stabilized \cite{NRSS91,SSNR91,PhysRevB.44.2664} by condensing a Higgs scalar with U(1) charge 2. We will meet both mechanisms in the model of Section~\ref{sec:lattice}.
The resulting state has co-existing topological order and spontaneous charge currents.

\section{SU(2) lattice gauge theory}
\label{sec:lattice}

This section will extend the SU(2) gauge theory of Refs.~\onlinecite{SS09,DCSS15b,DCSS16,2016PhRvB..94k5147S} 
to obtain lattice model realizations of the 
physics sketched in Section~\ref{sec:o3}. The SU(2) gauge theory was initially proposed as a convenient reformulation of 
particular theories of topological order in insulators \cite{NRSS89,NRSS90,NRSS91,SSNR91} and 
metals \cite{2007PhRvB..75w5122K,2008NatPh...4...28K,2008PhRvB..78d5110K}, which also allowed one to recover the large Fermi surface Fermi liquid at large doping. For our purposes, it also turns out to be a convenient setting in which to realize the states discussed in Section~\ref{sec:o3}.
The theory explicitly includes charged fermionic excitations, and so it is possible to obtain a gap near the antinodes,
and also directly compute the charged currents.

We start with electrons $c_{i \alpha}$ on the square lattice with dispersion
\beq
\mathcal{H}_c = - \sum_{i,\rho} t_\rho \left( c_{i , \alpha}^\dagger c_{i + {\bm v}_\rho, \alpha}^{\vphantom \dagger}  + 
c_{i + {\bm v}_\rho, \alpha}^\dagger c_{i, \alpha}^{\vphantom \dagger} \right)  -\mu \sum_{i} c_{i , \alpha}^\dagger c_{i , \alpha}^{\vphantom \dagger}  + \mathcal{H}_{\rm int} \label{Hc}
\eeq
 As discussed above Eq.~(\ref{bloch3}), we label {\it half} the links from site $i$ by the index $\rho$ and the vector ${\bm v}_\rho$: to avoid double-counting the vectors ${\bm v}_\rho$
 do not contain any pair that add to 0. With first, second, and third neighbors, ${\bm v}_\rho$ ranges over the 6 vectors
 $\hat{\bm x}$, $\hat{\bm y}$, $\hat{\bm x}+\hat{\bm y}$, $-\hat{\bm x} + \hat{\bm y}$, $2 \hat{\bm x}$, and $2 \hat{\bm y}$.

We represent the interactions between the electrons in a `spin-fermion' form \cite{hertz} 
using an on-site field $\Phi^\ell (i)$, $\ell = x, y, z$, which is conjugate to the spin moment
on site $i$:
\beq
\mathcal{H}_{\rm int} = - \lambda \sum_i \Phi^\ell (i) c_{i , \alpha}^\dagger \sigma^\ell_{\alpha\beta} c_{i , \beta}^{\vphantom \dagger} + V_\Phi \label{Hint}
\eeq
where $\sigma^\ell$ are the Pauli matrices. We leave the effective action for $\Phi$ in $V_\Phi$ unspecified - different choices for $V_\Phi$ allow
us to tune between the phases discussed below.

The key to obtaining insulators and metals with topological order (and hence a pseudogap without breaking translational symmetry)
is to transform the electrons to a rotating reference frame \cite{SS09,DCSS15b,DCSS16,2016PhRvB..94k5147S}
along the local magnetic order, using a SU(2) rotation $R_i$ and (spinless-)fermions 
$\psi_{i,s}$ with $s= \pm$,
\beq
\left( \begin{array}{c} c_{i\uparrow} \\ c_{i\downarrow} \end{array} \right) = R_i \left( \begin{array}{c} \psi_{i,+} \\ \psi_{i,-} \end{array} \right),
\label{R}
\eeq
where 
\beq
R_i^\dagger R_i = R_i R_i^\dagger = 1. 
\eeq
Note that this representation immediately introduces a SU(2) gauge invariance (distinct from the global SU(2) spin rotation)
\bea
\left( \begin{array}{c} \psi_{i,+} \\ \psi_{i,-} \end{array} \right) &\rightarrow& V_i \left( \begin{array}{c} \psi_{i,+} \\ \psi_{i,-} \end{array} \right) \label{gauge1} \\
\quad R_i &\rightarrow R_i& V_i^\dagger , \label{gauge2}
\eea
under which the original electronic operators remain invariant, $c_{i\alpha}\rightarrow c_{i\alpha}$; here $V_i (\tau) $ is a  SU(2) gauge-transformation acting on the $s=\pm$ index. So the $\psi_s$ fermions are SU(2) gauge fundamentals, carrying 
the physical electromagnetic global U(1) charge, but not the SU(2) spin of the electron: they are the fermionic  
``chargons'' of this theory, and  the density of the $\psi_s$ is the same as that of the electrons.
The bosonic $R$ fields also carry the global SU(2) spin (corresponding to left multiplication of $R$) but are electrically neutral:
they are the bosonic ``spinons''. We will relate them below to the spinons, $z_\alpha$, of the $\mathbb{CP}^1$ model in Eq.~(\ref{Sz}).
A useful summary of the gauge and global symmetry quantum numbers of the various fields is in Table~\ref{tab:charge}.
\begin{table}
\begin{center}
\begin{tabular}{|c|c|c|c|c|c|}
\hline
Field & Symbol & Statistics & SU(2)$_{\rm gauge}$ & SU(2)$_{\rm spin}$ & U(1)$_{\rm e.m. charge}$ \\
\hline 
Electron & $c$ & fermion & ${\bm 1}$ & ${\bm 2}$ & -1\\
Spin magnetic moment & $\Phi$ & boson & ${\bm 1}$ & ${\bm 3}$ & 0 \\
Chargon & $\psi$ & fermion & ${\bm 2}$ & ${\bm 1}$ & -1 \\
Spinon & $R$ or $z$ & boson & $\bar{\bm 2}$ & $ {\bm 2}$ & 0 \\
Higgs & $H$ & boson & ${\bm 3}$ & ${\bm 1}$ & 0 \\
\hline
\end{tabular}~\\~\\
\end{center}
\caption{Quantum numbers of the matter fields in the SU(2) Lattice gauge theory.
The transformations under the SU(2)'s are labelled by the dimension
of the SU(2) representation, while those under the electromagnetic U(1) are labeled by the U(1) charge.
The spin correlations are characterized by $\Phi$ in Eq.~(\ref{Hint}).
The Higgs field is the transform of $\Phi$ into a rotating reference frame via Eq.~(\ref{e2}).}
\label{tab:charge}
\end{table}

\begin{figure}
\begin{center}
\includegraphics[height=5in]{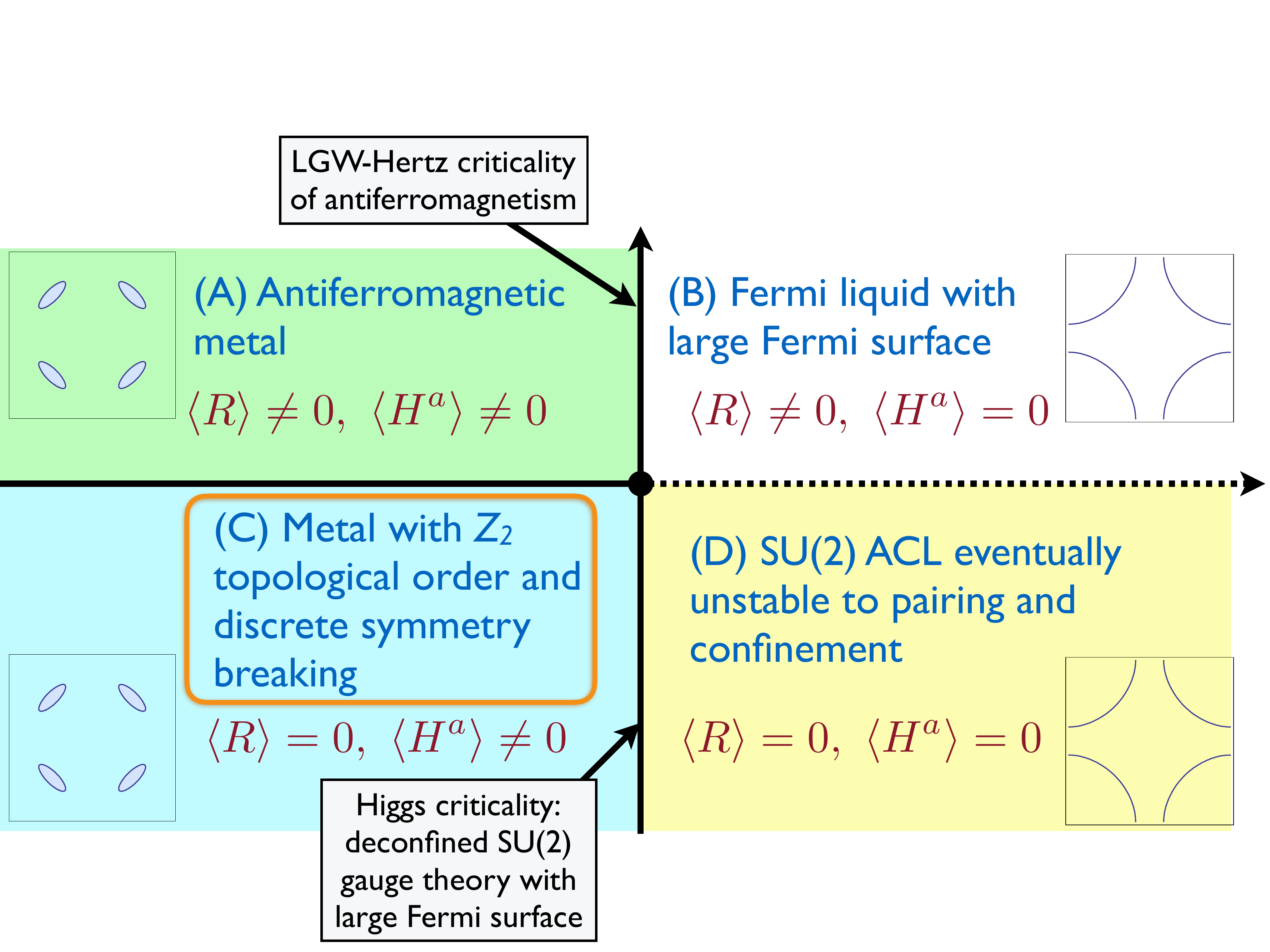}
\end{center}
\caption{Phase diagram of the SU(2) lattice gauge theory adapted from Ref.~\onlinecite{DCSS16}. The $x$ and $y$ axes are parameters controlling the condensates of $H$ and $R$ respectively. There is long-range antiferromagnetic order only in phase A. The Landau-Ginzburg-Wilson-Hertz theory \cite{hertz} describing transition between the conventional phases A and B is believed to provide a suitable framework for the Fe-based superconductors \cite{RMFAC17}. The hole-doped cuprate superconductors are proposed to follow the route A-C-D-B with increasing doping. Our interest here is in the pseudogap metal phase C. The optimal doping criticality \cite{LTCP15} is the transition from C to D, where the Higgs condensate vanishes in the presence of a large Fermi surface of fermions carrying SU(2) gauge charges. Phase D describes the overdoped regime, and is proposed to underlie the extended regime of criticality found in a 
magnetic field \cite{Cooper603}, and the non-BCS superconductivity \cite{Bozovic}.    }
\label{fig:phasediag}
\end{figure}

Inserting the parameterization in Eq.~(\ref{R}) into $\mathcal{H}_{\rm int}$, we can write Eq.~(\ref{Hint}) as
\beq
\mathcal{H}_{\rm int} = - \lambda \sum_i  H^a (i) \, \psi_{i ,s}^\dagger \, \sigma^a_{ss'} \, \psi_{i  ,s'}^{\vphantom \dagger}  + V_H \label{Hint2}
\eeq
We have introduced here the on-site Higgs
field $H^a (i)$, where $a=x,y,z$ and $\sigma^a$ are the Pauli matrices. This is the spin magnetic moment transformed
into the rotating reference frame, and is related to $\Phi^\ell (i)$ via
\beq
H^a (i) = \frac{1}{2} \Phi^\ell (i) \mbox{Tr} \left[ \sigma^\ell R_i^{\vphantom \dagger} \, \sigma^a R_i^\dagger \right], \label{e2}
\eeq
and the inverse relation
\beq
\Phi^\ell (i)= \frac{1}{2}  H^a (i)  \mbox{Tr} \left[ \sigma^\ell R_i^{\vphantom \dagger} \, \sigma^a R_i^\dagger \right]. \label{e2a}
\eeq
These relations can also be written as
\beq 
\sigma^a H^a (i) =   R_i^\dagger \, \sigma^\ell \Phi^{\ell} (i) \, R_i^{\vphantom \dagger}  . \label{e2b}
\eeq
The Higgs field transforms as an adjoint under the SU(2) gauge transformation, but does not carry spin or charge (see Table~\ref{tab:charge})
\beq
H^a (i) \rightarrow \frac{1}{2} H^b (i) \mbox{Tr} \left[ \sigma^a V_i \, \sigma^b V_i^\dagger  \right], \label{Hgauge}
\eeq
or equivalently
\beq 
\sigma^a H^a (i) \rightarrow   V_i \, \sigma^b H^b (i) \, V_i^\dagger  . \label{Hgauge2}
\eeq
We recall in Fig.~\ref{fig:phasediag} an earlier mean-field 
phase diagram \cite{DCSS16} obtained by condensing $R$ or $H$ or both. Our interest here will be primarily in phase C, which
has $\mathbb{Z}_2$ topological order because the condensation of the Higgs field breaks the SU(2) invariance down to $\mathbb{Z}_2$. 

We focus here on the  
effective Hamiltonian for the `chargons', the $\psi$ fermions in phase C. This is motivated by our aim of eventually
computing the charge currents. To obtain the Hamiltonian, we insert the parameterization in Eq.~(\ref{R}) into the hopping terms in 
$\mathcal{H}_c$, and decouple the resulting quartic terms.   Such an effective Hamiltonian has the form
\beq
\mathcal{H}_\psi =  - \sum_{i,\rho}\left( w_\rho  \psi_{i ,s}^\dagger \, U_{ss'}^\rho (i) \, \psi_{i + {\bm v}_\rho, s'}^{\vphantom \dagger}  + \mbox{H.c.} \right)
-\lambda \sum_i H^a (i) \, \psi_{i ,s}^\dagger \, \sigma^a_{ss'} \, \psi_{i  ,s'}^{\vphantom \dagger} - \mu
\sum_i  \psi_{i ,s}^\dagger  \psi_{i  ,s}^{\vphantom \dagger} \label{Hpsi}
\eeq
The magnitudes of the bare hoppings of the $\psi$ fermions are determined by the real numbers $w_\rho$; for simplicity, we fix
these hopping parameters at their bare values $w_\rho = t_\rho$. We have also included a SU(2) matrix on every link, $U^\rho (i)$, 
which represents the gauge connection used by the $\psi$ fermions to hop between sites. This clearly transforms under the gauge
transformation in Eq.~(\ref{gauge1},\ref{gauge2}) via
\beq
U^\rho (i) \rightarrow V_{i}^{\vphantom \dagger} U^\rho (i) V_{i + {\bm v}_\rho}^\dagger . \label{Ugauge}
\eeq
The previous analyses of this model \cite{SS09,DCSS15b,DCSS16,2016PhRvB..94k5147S} only examined the unit SU(2) matrix case $U^\rho = \mathbb{I}$. Below, we will describe other choices for $U^\rho$, and show that they can lead to states with spontaneous charge currents: 
this is the main new proposal in this paper for the SU(2) lattice gauge theory.

We will work with a translationally invariant ansatz \cite{Wen2002} 
for the SU(2) gauge-charged fields, $U^\rho (i)$ and $H^a (i)$, which can be taken to be independent of $i$. However, to make contact with 
earlier formulations in which the SU(2) is broken down to a U(1) or $\mathbb{Z}_2$ gauge theory \cite{SS09,DCSS15b,DCSS16}, 
it is useful to sometimes perform a gauge transformation to a spatially-dependent ansatz. 
The spatially dependent form cannot be gauge transformed back to the translationally invariant form using only the U(1) or $\mathbb{Z}_2$ gauge transformations, and so the spatial dependence is not optional
in the U(1) or $\mathbb{Z}_2$ gauge theories. 
We choose the space dependence of the SU(2) gauge fields
in the following form
\bea
U^\rho (i) &=&  V_{i}^{\vphantom \dagger} \left[ \exp \left( i \theta_\rho \ell_\rho^{ a} \sigma^a \right) \exp \left(  -\frac{i}{2} ({\bm Q} \cdot {\bm v}_{\rho}) \sigma^z \right) \right]
V_{i + {\bm v}_\rho}^\dagger  \nn
\sigma^a H^a (i) &=& V_{i}^{\vphantom \dagger} \, \sigma^b \Theta^b \,  V_{i}^{\dagger},
 \label{ansatz}
\eea
where
\beq
V_i =  \exp \left( - \frac{i}{2} ({\bm Q} \cdot {\bm r}_i) \sigma^z \right).
\eeq
The background gauge and Higgs fields are fully 
specified by the wavevector ${\bm Q}$, the 3 real numbers $\Theta^a$, the angle $\theta_\rho$ and the unit vector $\ell^a_\rho$
($\sum_{a = x,y,z} (\ell^{a}_\rho)^2 = 1$) on each $\rho$ link. Note that the ${\bm r}_i$ dependence is purely in fields performing the gauge transformation
so all gauge-invariant combinations will be translationally invariant. 
In component form, we can write Eq.~(\ref{ansatz}) as
\bea
H^x (i) \pm i H^y (i) &=& (\Theta^x \pm i \Theta^y) e^{\pm i {\bm Q} \cdot {\bm r}_i} \nn\
H^z (i) &=& \Theta^z \nn
U^\rho (i) &=& \cos(\theta_\rho) + i \sin (\theta_\rho) \Bigl[ 
\ell_\rho^z \sigma^z + (\ell_\rho^x - i \ell_\rho^y) e^{-i {\bm Q} \cdot {\bm r}_i} \sigma^+ + (\ell_\rho^x + i \ell_\rho^y) e^{i {\bm Q} \cdot {\bm r}_i} \sigma^- \Bigr] \label{ansatz2}
\eea
where $\sigma^+ = (\sigma^x + i \sigma^y)/2$.

The remaining task before us is to describe the physical properties of the phases obtained for different values of the parameters
$\Theta^a$, $\theta_\rho$, $\ell^a_\rho$ which are determined by
minimizing a suitable free energy. We will do this first for the previously studied phases  in Section~\ref{sec:previous}, and then in Section~\ref{sec:loop} for the new phases obtained here. We will find that almost all ansatzes break the SU(2)
gauge symmetry to a smaller gauge group: this Higgs phenomenon is accompanied by the appearance
of topological order and the gapping of the fermionic spectrum to yield a pseudogap state.

Before turning to this task, we note the transformations of the $\psi_i$, $R_i$, $H^a (i)$ and $U^\rho (i)$
under symmetries in Table~\ref{table:o3}. The simplest choice is to 
assign the transformations 
so that they commute with SU(2) gauge transformations. Then the transformations under spatial
symmetries ($T_x$, $T_y$, $I_x$, and $I_y$) are equal to the identity in SU(2) space, 
and simply given by the transformations on the spatial indices. More non-trivial are the 
transformations under time reversal, $\mathcal{T}$; these we assign as
\begin{equation}
\mathcal{T}: \psi \rightarrow - i \sigma^y \, \psi \quad,\quad R \rightarrow R \quad , \quad
H \rightarrow - H \quad, \quad U \rightarrow U,
\label{timesu2}
\end{equation}
along with the anti-unitary complex conjugation.

We close this discussion by pausing to recall 
the reasoning \cite{1994PhRvL..72.2089C,2003RvMP...75..913S,2016PhRvB..94k5147S} for the presence of 
$\mathbb{Z}_2$ topological order in the Higgs state C in which the SU(2) gauge invariance has broken down to $\mathbb{Z}_2$, and
why such a state can have small Fermi pockets and a pseudogap even in the presence of translational symmetry.
To break SU(2) down to $\mathbb{Z}_2$, the configuration
of Higgs and link fields, $\Theta^a$ and $\ell_\rho^a$, must
transform under global SU(2) transformations like a SO(3) order parameter. Because $\pi_1 (SO(3)) = \mathbb{Z}_2$,
there are vortex line defects with single-valued Higgs 
and link fields. Such a defect must also correspond to a single-valued vortex configuration of the antiferromagnetic order. Now we imagine undoing
the vortex configuration by choosing
$R$ such that the $\psi$ fermions observe a locally constant background
in $\mathcal{H}_\psi$. Then we will find that $R$ is double-valued, with $R \rightarrow - R$ upon
encircling a loop around the vortex. Consequently, the $\psi$ fermions acquire a Berry phase of $\pi$ around the vortex, and the $\psi$ fermions and vortex excitations (the `visons' \cite{SenthilFisher}) are relative semions. 
These vortices will be suppressed in the Higgs-condensed ground state, 
and in such a ground state we can globally transform to a rotating reference frame in which
the $\psi$ fermions are described by $\mathcal{H}_\psi$. The ${\bm Q}$ dependent configuration of Higgs and link fields in Eq.~(\ref{ansatz2})
can then reconstruct the $\psi$ Fermi surface into pockets.

\subsection{Previously studied phases}
\label{sec:previous}

\subsubsection{Insulators with N\'eel or VBS order}
\label{sec:neel}

These are obtained from the saddle point with ${\bm Q} = (\pi, \pi)$ and 
$\Theta^a = (\Theta, 0, 0)$, while all the $\theta_\rho = 0$ so that $U^\rho = \mathbb{I}$. The Higgs field has two sublattice
order polarized the $x$ direction with $H^a (i) = \eta_i (\Theta, 0, 0)$ where $\eta_i = \pm 1$ on the two sublattices. 

The dispersion of
the $\psi$ fermions is the same as that of electrons in the presence of N\'eel order, and we obtain
the needed fermionic gap in the anti-nodal regions of the Brillouin zone.
Note however that Eq.~(\ref{e2a}) implies 
that the appearance of physical N\'eel order requires the condensation of $R$. We assume $\Theta$ is large, and choose
the chemical potential to lie within the band gap which has magnitude $|\Theta|$. Consequently, the $\psi$ fermions form a band insulator, and the charge gap is of order $|\Theta|$ which we assume
is of order the $U$ of the underlying Hubbard model.

We now argue that fluctuations about this `band insulator' 
saddle point are described by the $\mathbb{CP}^1$ model of Eq.~(\ref{Sz}). 
A key observation is that presence of the Higgs condensate $H^a (i) = \eta_i (\Theta, 0, 0)$ breaks the SU(2) gauge invariance
down to U(1). Such a Higgs condensate is invariant under residual U(1) gauge transformations about the $x$ axis. 
So we parameterize the the fluctuations of the link fields by
\beq
U^\rho = \exp \left( i \sigma^x {\bm a} \cdot v_{{\bm \rho}} \right); \label{UCP1}
\eeq
then ${\bm a}$ transforms like the spatial component of a U(1) gauge field under the residual gauge transformation. 
To obtain the spinons $z_\alpha$ in Eq.~(\ref{Sz}), we need to parameterize $R$ in terms $z_\alpha$ so that Eq.~(\ref{gauge2})
implies that $z_\alpha$ have unit gauge charge under the gauge transformation $V = \exp (i \sigma^x \zeta)$, where $\zeta$ generates
the gauge transformation. This is obtained from
\beq
R = \frac{1}{\sqrt{2}} 
\left( \begin{array}{ccc} z_\uparrow + z_\downarrow^\ast & & -z_\downarrow^\ast + z_\uparrow \\
z_\downarrow - z_\uparrow^\ast & & z_\uparrow^\ast + z_\downarrow
\end{array} \right), \label{Rz}
\eeq
under which $z_\alpha \rightarrow z_\alpha e^{-i \zeta}$.

We note here a subtlety in identifying the $z_\alpha$ and ${\bm a}$ above with the fields
of the $\mathbb{CP}^1$ model of Eq.~(\ref{Sz}): the symmetry assignments discussed near Eq.~(\ref{timesu2}) for the SU(2) gauge theory do not map under Eq.~(\ref{UCP1}) to the symmetry assignments in
Table~\ref{table:o3} and Ref.~\onlinecite{2007AnPhy.322.2635B}. The difference is 
present for transformations $T_x$, $T_y$ and $\mathcal{T}$, under which the Higgs field $\Theta \rightarrow -\Theta$ in the SU(2) formulation for ${\bm Q} = (\pi, \pi)$. 
In the $\mathbb{CP}^1$ formulation,
it is implicitly assumed that the Higgs field is invariant under all transformations. 
To remedy this, we need to combine 
the SU(2) gauge transformation $V = \exp(-i (\pi/2) \sigma^z)$ with the operations of 
$T_x$, $T_y$, and $\mathcal{T}$ in the SU(2) gauge theory.

Beyond the fluctuations described by the $\mathbb{CP}^1$ model, we have to consider the non-perturbative role of monopoles 
in the U(1) gauge field \cite{NRSS89,NRSS90}. In the earlier works, the spin liquid was described using Schwinger bosons with
a unit boson density per site. In the presence of monopoles, this background density of bosons contributed a net Berry phase \cite{NRSS90}. In the present formulation, we have a background of a filled band of the $\psi$ fermions. The monopole Berry phase computation of
Ref.~\onlinecite{NRSS90} (Section III.A) carries over with little change to the fermion case, and we obtain the same monopole Berry phases.

The remaining analysis of the $\mathbb{CP}^1$ model augmented with monopole Berry phases is as before \cite{NRSS89,NRSS90,senthil1,senthil2}. The phase with $\langle z_\alpha \rangle \neq 0$ has N\'eel order,
while the strong coupling phase $\langle z_\alpha \rangle = 0$ is initially a U(1) spin liquid which eventually
confines at the longest scales to a VBS; the transition between these phase is described by the critical $\mathbb{CP}^1$ model.

\subsubsection{Insulators with spiral spin order or $\mathbb{Z}_2$ topological order}
\label{sec:spiral}

The saddle point is similar to that in Section~\ref{sec:neel}, except that ${\bm Q}$ is incommensurate. 
So we have $\Theta^a = (\Theta, 0, 0)$, while all the $\theta_\rho = 0$ so that $U^\rho = \mathbb{I}$.
Eq.~(\ref{ansatz}) implies that the spatial dependence of the Higgs field is specified by
\beq
H^a (i) = \Theta ( \cos({\bm Q} \cdot {\bm r}_i ),  \sin({\bm Q} \cdot {\bm r}_i ), 0). \label{Hspiral}
\eeq
For generic ${\bm Q}$, there is no residual U(1) gauge invariance left by such a condensate. Instead,
the only residual gauge invariance is $\mathbb{Z}_2$, associated with the choice $V_i = \pm 1$. Consequently, the spin liquid
described by this Higgs condensate has $\mathbb{Z}_2$ topological order. Again, to obtain an insulator we assume that the 
chemical potential is within the gap of the $\psi$ bands.

The phases obtain by Eq.~(\ref{Hspiral}) are precisely those described in Refs.~\onlinecite{NRSS91,SSNR91,2016PhRvB..94b4502C}, 
and the earlier
analyses can be applied directly here. The phase with $R$ condensed has spiral spin order, while the phase with $R$ gapped
is a $\mathbb{Z}_2$ spin liquid. 

\subsubsection{Metals with topological order}
\label{sec:metaltopo}

A key advantage of the present SU(2) gauge theory formulation is that the results obtained in Sections~\ref{sec:neel}
and~\ref{sec:spiral} are immediately generalized from insulators to metals. One only has to change the chemical potential $\mu$
so that one of the $\psi$ bands is partially occupied, and we obtain a Fermi surface of $\psi$ chargons. 

For the $U(1)$ gauge theory in Section~\ref{sec:neel}, the $\psi$ Fermi surface can suppress the monopoles, and 
the U(1) topological order survives in an `algebraic charge liquid' (ACL) \cite{2008NatPh...4...28K}. The $\mathbb{Z}_2$ topological order
was already stable in the insulator in Section~\ref{sec:spiral}, and it continues to survive in the presence of the $\psi$ 
Fermi surface.

It is also possible that the ACL becomes a `fractionalized Fermi liquid' (FL*) \cite{FFL,TSMVSS04,APAV04}. 
This appears when the $\psi$ fermions
bind with the $R$ spinons to form `small' Fermi surfaces of 
electron-like quasiparticles \cite{2007PhRvB..75w5122K,2008NatPh...4...28K,2012PhRvB..85s5123P,2015PNAS..112.9552P} while retaining the topological order.

\subsection{States with SU(2) gauge fields on links}
\label{sec:loop}

We turn to our new results on the SU(2) lattice gauge theory. We will examine saddle points with 
non-zero Higgs condensate $\langle H \rangle \neq 0$ (as above) and also a non-trivial background gauge flux
$U^\rho \neq \mathbb{I}$.
We will find that such saddle points can break time-reversal and inversion symmetries in gauge-invariant observables,
and that is sufficient to induce charge currents. 
Ising-nematic order can also be present, as found previously, but it
can also co-exist with spontaneous charge currents. This subsection will report results in the
gauge ${\bm Q} = 0$. Recall that the value of ${\bm Q}$ is merely a gauge choice in the full SU(2) gauge theory (but not in U(1) or $\mathbb{Z}_2$ gauge theory formulations).
In this gauge, the Higgs field is $i$ independent 
with $H^a (i) = \Theta^a$.

Formally, we should integrate out the fermions in $\mathcal{H}_\psi$ in Eq.~(\ref{Hpsi}), and then minimize the 
resulting action functional for the Higgs and gauge fields. However, this is computationally demanding, and the structure assumed in $\mathcal{H}_\psi$ is phenomenological anyway. So we will be satisfied by minimizing a phenomenological gauge-invariant
functional of the Higgs and gauge fields, consisting of short-range terms that can be constructed out of a 
single plaquette. In metallic states, the fermion determinant can also induce longer-range terms with a power-law
decay, but we will not include those here: in our simple treatment, we assume that the dominant energy arises from the short-range terms.

The effective potential also has terms contributing to a Higgs potential $V_H$ which arise from $V_\Phi$ in Eq.~(\ref{Hint}) via Eq.~(\ref{e2a}). As we will not specify $V_H$, we assume that this potential has already 
been minimized to yield the values of $\Theta^a$. So we
will only consider the remaining free 
energy, $\mathcal{F}$, which is a function only of the $U^\rho$. 

The following gauge-invariant link variables are useful ingredients in constructing the free energy 
\begin{equation}
    \mathcal{L}^\rho = \Theta^a \Theta^b \, \mbox{Tr} (\sigma^a U^\rho \sigma^b U^{\rho\dagger} ) ;
\end{equation}
These link variables are even under the time-reversal operation described in Eq.~(\ref{timesu2}). In terms 
of these link variables, we can define the nematic order parameters in Eq.~(\ref{defnematic})
by
\beq 
\mathcal{N}_1 = \mathcal{L}^1 - \mathcal{L}^2 \quad , \quad \mathcal{N}_2 = \mathcal{L}^3
- \mathcal{L}^4. \label{nematicU}
\eeq  
In writing the free energy, it is useful to change notation and write the link variables via 
\beq 
U^\rho (i) \rightarrow U_{ij} \quad, \mbox{with}~{\bm r}_j = {\bm r}_i + {\bm v}_\rho. \label{latticeij}
\eeq
We minimized the free energy
\bea
\mathcal{F} &=& K_1 \left(\mathcal{L}^1 + \mathcal{L}^2\right)
+ K_8 (\mathcal{L}^1 - \mathcal{L}^2) ^2 + K_9 \left( \mathcal{L}^3
- \mathcal{L}^4\right)^2
\nn 
&+&
\sum_{\setlength\unitlength{0.5pt}
\begin{picture}(35,40)
\put(12,4){\line(1,0){20}}
\put(12,4){\line(0,1){20}}
\put(12,24){\line(1,-1){20}}
\put(34,0){\small $i$}
\put(0,0){\small $j$}
\put(6,27){\small $k$}
\end{picture}}
\Bigl[
K_3 \, \mbox{Tr} \left( U_{ij} U_{jk} U_{ki} \right) + K_4 \, \left[\mbox{Tr} \left( U_{ij} U_{jk} U_{ki} \right)\right]^2 + K_5 \, \mbox{Tr} \left( U_{ij} U_{jk} \sigma^a U_{ki} \sigma^b \right)
\Theta^a  \Theta^b \nn
&~&~~~- K_6 \, \mbox{Tr} \left( U_{ij} \sigma^a U_{jk}  U_{ki} \sigma^b \right)
\Theta^a  \Theta^b   \Bigr] +
\sum_{\setlength\unitlength{0.5pt}
\begin{picture}(35,40)
\put(12,4){\line(1,0){20}}
\put(12,4){\line(0,1){20}}
\put(12,24){\line(1,0){20}}
\put(32,4){\line(0,1){20}}
\put(34,0){\small $i$}
\put(0,0){\small $j$}
\put(0,27){\small $k$}
\put(31,27){\small $\ell$}
\end{picture}}~
 K_7 \, \mbox{Tr} \left( U_{ij} U_{jk} U_{k\ell} U_{\ell i} \right).
 \label{FU}
\eea
In the above expressions we assume that all terms obtained from the pictured symbols by square lattice symmetry 
operations have been summed over. This free energy depends upon 9 parameters $K_{1-9}$, and a priori they are
free to take arbitrary values. We used the residual SU(2)
gauge degree for freedom to set $\Theta^a = (\Theta, 0, 0)$, and then with 4 possible  values of the link variable
$\rho$, the free depends upon 12 real numbers which determine the $U^\rho$.

We characterized the free energy minima by their values 
of the nematic order parameters $\mathcal{N}_1$ and $\mathcal{N}_2$. We also need gauge-invariant observables
which are odd under time-reversal; for this we evaluated 
the combinations defined on right triangles, $ijk$:
\beq
\mathcal{P}_{ijk} = 
i \, \mbox{Tr} \left( \sigma^a U_{ij} U_{jk} U_{ki} \right)
H^a (i) .
\eeq
The spatial patterns of the $\mathcal{P}_{ijk}$, along
with the values of $\mbox{Tr} \left( U_{ij} U_{jk} U_{ki} \right)$, yield much information on the nature of time-reversal
and inversion symmetry breaking. Note that the $\mathcal{P}_{ijk}$
are non-zero only if the Higgs field is non-zero---this is a consequence of the transformation in Eq.~(\ref{timesu2}). So time-reversal symmetry can only be broken in states in which the
SU(2) gauge invariance is also broken.

Another important characterization of the states is provided by the values of the physical charge current. We used the values of the link variables obtained by the minimization of $\mathcal{F}$, and inserted them into the Hamiltonian $\mathcal{H}_\psi$ in 
Eq.~(\ref{Hpsi}).
We then determined the current on each link by evaluating the expectation value of the current operator
\beq
{\bm J}_\rho (i) =  - i {\bm v}_\rho \left( w_\rho  \psi_{i ,s}^\dagger \, U_{ss'}^\rho (i) \, \psi_{i + {\bm v}_\rho, s'}^{\vphantom \dagger}  - \mbox{H.c.} \right) \label{Jrho}
\eeq
in the fermion state specified by the Hamiltonian $\mathcal{H}_\psi$ at a low temperature. As shown in 
Appendix~\ref{app:mom}, for the background field configurations in Eq.~(\ref{ansatz2}), $\langle 
{\bm J}_\rho (i) \rangle$ turns
out to be independent of $i$ for general values of the variational parameters
$\Theta^a$, $\theta_\rho$, $\ell^a_\rho$ and ${\bm Q}$ in the Hamiltonian. This is as expected
from our arguments that Eq.~(\ref{ansatz2}) implies that all gauge-invariant observables should be translationally invariant. Moreover, we find that the value of $\langle {\bf J}_\rho \rangle$
always obeys Bloch's theorem in Eq.~(\ref{bloch3}); this is true in our numerics, 
and a general proof is in Appendix~\ref{app:mom}.

\begin{figure}
\begin{center}
\includegraphics[width=2.5in]{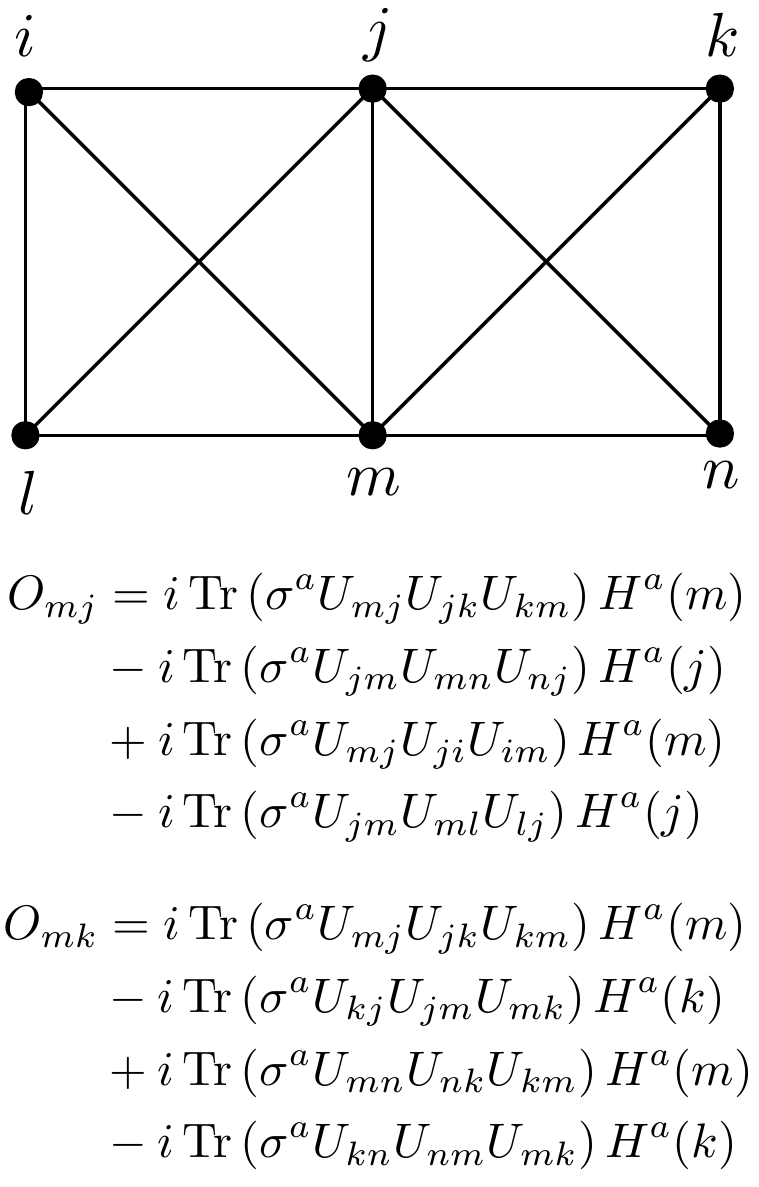}
\end{center}
\caption{Expressions for the time and inversion symmetry breaking order parameter ${\bm O}$
in terms of the variables of the SU(2) gauge theory. We use the same notation as in Eq.~(\ref{latticeij}) for the link values of $U^\rho (i)$ and ${\bm O}$ with $O_{ij} = {\bm O} \cdot {\bm v}_{\rho}$
when ${\bm r}_j = {\bm r}_i + {\bm v}_\rho$.}
\label{fig:O}
\end{figure}
It is also useful to examine local gauge-invariant operators
which have the same symmetry signatures as the physical current
${\bm J}_\rho$. Such operators will be realizations of the 
operator ${\bm O}$ characterizing states with broken
inversion and time-reversal symmetry. We obtained expressions using the symmetry
transformations described near Eq.~(\ref{timesu2}), 
and one set of operators is presented in Fig.~\ref{fig:O}. A derivation based upon a large $|H^a(i)|$
expansion is presented in Appendix~\ref{app:realspace}, along with other sets of possible operators.
For the translationally invariant solution and ${\bm Q}=0$ gauge being considered here, Fig.~\ref{fig:O} yields these expressions
for the order parameters $O_\rho$ along the directions ${\bm v}_\rho$:
\bea 
O_1 &=& i \, \mbox{Tr} \left(\sigma^a U^1 U^{2 \dagger} U^4 \right) \Theta^a - i \, \mbox{Tr} \left(\sigma^a U^{1 \dagger} U^{2 \dagger} U^{3} \right) \Theta^a +
 i \, \mbox{Tr} \left(\sigma^a U^{1} U^{2} U^{3 \dagger} \right) \Theta^a
- i \, \mbox{Tr} \left(\sigma^a U^{1 \dagger} U^{2} U^{4 \dagger} \right) \Theta^a  \nn
O_2 &=& i \, \mbox{Tr} \left(\sigma^a U^2 U^1 U^{3 \dagger} \right) \Theta^a 
- i \, \mbox{Tr} \left(\sigma^a U^{2 \dagger} U^1 U^{4} \right) \Theta^a + 
i \, \mbox{Tr} \left(\sigma^a U^2 U^{1\dagger} U^{4 \dagger} \right) \Theta^a
- i \, \mbox{Tr} \left(\sigma^a U^{2 \dagger} U^{1 \dagger} U^{3} \right) \Theta^a \nn
O_3 &=& i \, \mbox{Tr} \left(\sigma^a U^2 U^1 U^{3 \dagger} \right) \Theta^a 
- i \, \mbox{Tr} \left(\sigma^a U^{1 \dagger} U^{2 \dagger} U^{3} \right) \Theta^a + 
i \, \mbox{Tr} \left(\sigma^a U^1 U^{2} U^{3 \dagger} \right) \Theta^a
- i \, \mbox{Tr} \left(\sigma^a U^{2 \dagger} U^{1 \dagger} U^{3} \right) \Theta^a \nn 
O_4 &=& i \, \mbox{Tr} \left(\sigma^a U^2 U^{1 \dagger} U^{4 \dagger} \right) \Theta^a 
- i \, \mbox{Tr} \left(\sigma^a U^{1} U^{2 \dagger} U^{4} \right) \Theta^a + 
i \, \mbox{Tr} \left(\sigma^a U^{1 \dagger} U^{2} U^{4 \dagger} \right) \Theta^a
- i \, \mbox{Tr} \left(\sigma^a U^{2 \dagger} U^{1} U^{4} \right) \Theta^a \nn 
\eea 
These will be connected to ${\bm J}_\rho$ via an expression like Eq.~(\ref{JKO}). Note 
that the $O_\rho$ can only be non-zero
when the Higgs condensate is non-zero, because only the Higgs field
is odd under time-reversal in Eq.~(\ref{timesu2}). An explicit demonstration that a non-zero
charge current requires a non-zero Higgs field is in Appendix~\ref{app:Higgs}.

Turning to the minimization of $\mathcal{F}$ in Eq.~(\ref{FU}), we did not perform an exhaustive search of different classes of states over the 9 parameters,
$K_{1-9}$ in $\mathcal{F}$. Rather we explored a few values to yield representative minima,
and will describe a few of the typical states in the subsections below. All of the minimization
was performed with the Higgs field oriented along the $x$ direction, $\Theta^a = (\Theta, 0, 0)$.

\subsubsection{Symmetric state}

This state preserves all square lattice symmetries and time reversal. We obtained such a
minimum at $K_1=1$, $K_2 = -1$, $K_3 = 2$, $K_4 =2$, $K_5 =2$, $K_6 =2$, $K_7 = 0.1$,
$K_8=0$, $K_9=0$. In the ${\bm Q}=0$ gauge, the link fields take the values:
\bea
U^1 &=& i \sin (\theta) \, \sigma^x - i \cos (\theta) \, \sigma^y \nn 
U^2 &=& i \sin (\theta) \, \sigma^x + i \cos (\theta) \, \sigma^y \nn
U^3 &=& -1 \nn 
U^4 &=& 1,
\eea 
where $\theta = 0.344 \pi$. All the $O_\rho$, $\mathcal{N}_1$ and $\mathcal{N}_2$ order
parameters, and the currents ${\bm J}_\rho$, vanish in this state. The SU(2) gauge invariance is broken down to $\mathbb{Z}_2$ because the $U^\rho$ and $\Theta^a$ have no common orientation, 
and so this state has $\mathbb{Z}_2$ topological order.

\subsubsection{Ising-nematic order}
\label{sec:nematic}

This state preserves time reversal and inversion, but breaks a square lattice rotation symmetry. 
We obtained such a
minimum at $K_1=0.5$, $K_2 = 0.5$, $K_3 = -1$, $K_4 =0.25$, $K_5 =0$, $K_6 =0$, $K_7 = 0$,
$K_8=0$, $K_9=5$. In the ${\bm Q}=0$ gauge the link fields take the values:
\bea
U^1 &=& -i \, \sigma^z \nn 
U^2 &=& \cos (\theta_1) + i \sin (\theta_1) \,  \sigma^z \nn
U^3 &=& \cos (\theta_2) + i \sin (\theta_2) \,  \sigma^z \nn
U^4 &=& -\cos (\theta_2) - i \sin (\theta_2) \,  \sigma^z
\eea 
where $\theta_1 = 0.672 \pi$ and $\theta_2 = 0.427  \pi$. All the $O_\rho$, and the currents ${\bm J}_\rho$, vanish in this state. 
However the nematic order $\mathcal{N}_1 \neq 0$, while $\mathcal{N}_2 =0$.
Note that the $U^\rho$ are oriented along a common $z$ direction, while the Higgs field $\Theta^a$
is oriented along the distinct $x$ direction. So SU(2) gauge invariance is broken down $\mathbb{Z}_2$,
and $\mathbb{Z}_2$ topological order is present. In the insulator, 
this state has the same properties as the ``$(\pi, q)$ SRO'' state 
of
Refs.~\onlinecite{NRSS91,SSNR91}. 

\subsubsection{State with broken time-reversal}

Now we present a state which breaks time-reversal but {\it not\/} inversion. So this state has
no spontaneous currents, and $O^\rho =0 $ and ${\bm J}^\rho =0 $. Nevertheless, time reversal
is broken as signaled by the non-zero values of some of the $\mathcal{P}_{ijk}$. Roughly speaking,
such a state has spontaneous currents along different directions in the gauge group, but the
net electromagnetic current vanishes. We obtained such a
minimum at $K_1=0.5$, $K_2 = 0.5$, $K_3 = 1$, $K_4 =0.667$, $K_5 =0$, $K_6 =0$, $K_7 = 1$,
$K_8=5$, $K_9=5$. In the ${\bm Q}=0$ gauge the link fields take the values:
\bea
U^1 &=& -\cos (\theta_1) - i \sin (\theta_1) \,  \sigma^z \nn 
U^2 &=& \cos (\theta_1) + i \sin (\theta_1) \,  \sigma^y \nn
U^3 &=& \cos (\theta_2) + i \sin (\theta_2) \,  (\sigma^y + \sigma^z)/\sqrt{2} \nn
U^4 &=& -\cos (\theta_2) + i \sin (\theta_2) \,  (\sigma^y - \sigma^z)/\sqrt{2},
\eea 
where $\theta_1 = 0.446 \pi$ and $\theta_2 = 0.497  \pi$. Again, SU(2) gauge invariance is broken down $\mathbb{Z}_2$,
and $\mathbb{Z}_2$ topological order is present.

\subsubsection{States with spontaneous charge currents.}

Finally, we turn to a description of the states presented in Fig.~\ref{fig:currents2}.

First, we present a state with the symmetry of Fig.~\ref{fig:currents2}a.
Such a state was obtained for $K_1=0.5$, $K_2 = 0.5$, $K_3 = -1$, $K_4 = 0.25$, $K_5 =0$, $K_6 =0$, $K_7 = 0$,
$K_8=2$, $K_9=5$. In the ${\bm Q}=0$ gauge the link fields take the values
\bea
U^1 &=& \cos (\theta_1) + i \sin (\theta_1) \, (\cos( \phi_1) \sigma^x + \sin (\phi_1) \sigma^z) \nn
U^2 &=& i \, (\cos( \phi_2) \sigma^x + \sin (\phi_2) \sigma^z) \nn
U^3 &=& -\cos (\theta_2) + i \sin (\theta_2) \, (\cos( \phi_1) \sigma^x + \sin (\phi_1) \sigma^z) \nn
U^4 &=& \cos (\theta_2) + i \sin (\theta_2) \, (\cos( \phi_1) \sigma^x + \sin (\phi_1) \sigma^z) 
\eea 
where $\theta_1 = 0.410 \pi$, $\phi_1 = 0.5063 \pi$, $\phi_2 = 0.558 \pi$, $\theta_2 = 0.387 \pi$.
This state has the $O^\rho$ and ${\bm J}^\rho$ non-zero, along with a non-zero Ising-nematic
order $\mathcal{N}_1 \neq 0$, but $\mathcal{N}_2 = 0$. So it has the full generic symmetry structure
of Fig.~\ref{fig:currents2}a. 
The gauge field configuration shows that SU(2) is broken down to $\mathbb{Z}_2$, and so $\mathbb{Z}_2$ topological order is present.
States in this class were the most common in our search over the
parameters $K_{1-9}$ among those
that broke time-reversal symmetry.

Among states with a residual U(1) gauge invariance, we found global minima with the symmetry of Fig.~\ref{fig:currents2}a only when
we restricted the search to states in which the Higgs and link fields were collinear in the gauge SU(2) space. We can parameterize the fluctuations about such a saddle point by multiplying the $U^\rho$
by the factor in Eq.~(\ref{UCP1}), and then we obtain a theory of a gapless U(1) photon $a_\mu$.
Because of the presence of the breaking of inversion and time-reversal symmetries, this action
will take the form in Eq.~(\ref{Sa}), including the term proportional to ${\bm \Gamma}$. As in
Section~\ref{sec:neel}, we have to consider the non-perturbative effects of monopoles: 
such a state can be stable against monopole proliferation only in the presence
of gauge-charged Fermi surfaces.

Next, we present a state with the symmetry of Fig.~\ref{fig:currents2}b.
Such a state was obtained for $K_1=0.5$, $K_2 = 0.5$, $K_3 = 1$, $K_4 = 0.667$, $K_5 =0$, $K_6 =0$, $K_7 = 0$,
$K_8=5$, $K_9=-1$. In the ${\bm Q}=0$ gauge the link fields take the values
\bea
U^1 &=& -\cos (\theta_1) - i \sin (\theta_1) \, \sigma^z  \nn
U^2 &=& \cos (\theta_1) + i \sin (\theta_1) \, \sigma^z \nn 
U^3 &=& \cos (\theta_2) + i \sin (\theta_2) \, \sigma^z \nn 
U^4 &=& \cos (\theta_3) - i \sin (\theta_3) \, \sigma^x 
\eea
where $\theta_1 = 0.451 \pi$, $\theta_2 = 0.503 \pi$, $\theta_3 = 0.379 \pi$. 
The order parameters $O^\rho$ and the currents ${\bm J}^\rho$ are
non-zero, and are consistent with the pattern in Fig.~\ref{fig:currents2}a.
There is also a non-zero Ising-nematic
order $\mathcal{N}_2 \neq 0$, but $\mathcal{N}_1 = 0$. The non-collinear alignment of the
gauge and Higgs fields indicates the presence of $\mathbb{Z}_2$ topological order.

\section{Conclusions}
\label{sec:conc}

We have presented 
computations showing that emergent background gauge connections, and associated Berry phases, arising from 
the local antiferromagnetic spin correlations
can induce spontaneous charge currents, while preserving translational symmetry. The main requirement on the gauge
theory is that {\it gauge-invariant} observables break time-reversal and inversion, but preserve
translation. 
At the same time, the topological order associated with the emergent gauge fields can account for the
anti-nodal gap in the charged fermionic excitations. 

The specific model we used for a stable pseudogap metal
had $\mathbb{Z}_2$ topological order. We employed a SU(2) lattice gauge theory with a Higgs field to realize such a phase. Going beyond earlier work on this theory, we
allowed the SU(2) gauge fields on the links to acquire
non-trivial values in the saddle point of the Higgs phase. These link fields had two important consequences. 
First, it became possible to obtain $\mathbb{Z}_2$
topological order even under conditions in which the proximate magnetically
ordered phase had collinear spin correlations at $(\pi, \pi)$; earlier 
realizations \cite{NRSS91,SSNR91,SS09,DCSS15b,DCSS16} required non-collinear spiral spin correlations. Second, the gauge-invariant combinations of the SU(2) gauge fluxes and the Higgs field could
break time-reversal and inversion symmetries without breaking translational symmetry. This allowed the appearance of 
spontaneous charge currents and Ising-nematic order in the Higgs phase. Linking the discrete broken symmetries to the presence of the Higgs condensate also explains why the broken symmetries do not survive in the confining phases at large doping.

An attractive features of our results is that
the charge currents, and the anti-nodal gap,
continue largely unmodified across transitions to states with long-range antiferromagnetic order, but without topological order. This is consistent with recent experiments \cite{2016NatPh..12...32Z,2017arXiv170106485J} on Sr$_2$Ir$_{1-x}$Rh$_x$O$_4$ showing
co-existence of N\'eel order and inversion and time reversal breaking over a certain range of doping and temperature.

The possible patterns of symmetry breaking in the translationally-invariant states with broken time-reversal and inversion
(but not their product) are illustrated in Fig.~\ref{fig:currents2}. Both states also break a lattice
rotation symmetry, and so they also have Ising-nematic order. The state in Fig.~\ref{fig:currents2}b
has the same pattern of symmetry breaking as states considered earlier \cite{2002PhRvL..89x7003S,2004PhRvB..69x5104S,2008PhRvL.100b7003B}. However, the state in Fig.~\ref{fig:currents2}a does not
appear to have been described previously in the literature. The Fig.~\ref{fig:currents2}a state has the attractive feature that its
Ising-nematic order is precisely that observed in other experiments \cite{2002PhRvL..88m7005A,Hinkov597,2010Natur.463..519D,2010Natur.466..347L}. 
The onset of Ising-nematic order and time-reversal and inversion symmetry breaking
could happen at the same or distinct temperatures, as we also found states in Section~\ref{sec:nematic} with Ising-nematic order but no charge currents. However, if a particular symmetry is broken in the pseudogap phase
(phase C in Fig.~\ref{fig:phasediag}), it must be restored when the Higgs condensate vanishes in the over-doped regime (phases D and B in Fig.~\ref{fig:phasediag}).

The existing experiments \cite{2016NatPh..12...32Z,2017arXiv170106485J} do not contain the polarization analysis needed to distinguish
between the states in Fig.~\ref{fig:currents2}a and b, and we hope such experiments will be undertaken.

We placed our results in the context of a global phase diagram for antiferromagnetism and superconductivity in two dimensions in Fig.~\ref{fig:phasediag}. In particular, we noted that this
phase diagram \cite{SS09,DCSS15b,DCSS16} is in accord with experiments exploring 
the hole-doped cuprates over a range of carrier density. 
Badoux {\it et al.\/} \cite{LTCP15} observe pseudogap metal at low $T$ and large doping, without
any charge density wave order: this is a candidate for our phase C. We note recent theoretical works \cite{2016PhRvL.117r7001E,CSE17} which studied electrical and thermal
transport across the phase transition C $\rightarrow$ D in Fig.~\ref{fig:phasediag},
and found results in good accord with observations \cite{LTCP15,Laliberte2016-arXiv,Collignon2016-arXiv,Michon2017}.
Cooper {\it et al.\/} \cite{Cooper603} observe an extended overdoped regime of linear-in-$T$
resistivity when the superconductivity is suppressed by a magnetic field: our phase D could be
such a critical phase. And Bo{\v z}ovi{\'c} {\em et al.\/} \cite{Bozovic} see strong deviations
from BCS theory in the doping and temperature dependence of the superfluid stiffness in the overdoped
regime: this could also be described by phase D.

\section*{Acknowledgements}

We thank E.~Berg, P.~Bourges, D.~Hsieh, S.~Kivelson, L.~Taillefer, C.~Varma, and A.~Vishwanath for valuable discussions.
This research was supported by the NSF under Grant DMR-1360789 and the MURI grant W911NF-14-1-0003 from ARO. Research at Perimeter Institute is supported by the Government of Canada through Industry Canada and by the Province of Ontario through the Ministry of Research and Innovation. SS also acknowledges support from Cenovus Energy at Perimeter Institute.

\appendix

\section{Momentum space}
\label{app:mom}

This appendix presents a few expressions from Section~\ref{sec:lattice} in momentum space. These expressions were used for our numerical 
computation.

The momentum space form of the electron dispersion in Eq.~(\ref{Hc}) is
\beq
\mathcal{H}_c = -  2\sum_{{\bm k},\rho} t_\rho  \cos( {\bm k} \cdot {\bm v}_\rho ) c_{{\bm k}, \alpha}^\dagger c_{{\bm k}, \alpha}^{\vphantom \dagger}
- \mu \sum_{{\bm k}} c_{{\bm k}, \alpha}^\dagger c_{{\bm k}, \alpha}^{\vphantom \dagger}  + \mathcal{H}_{\rm int}.
\eeq

The Hamiltonian for the $\psi$ fermions is obtained from Eqs.~(\ref{Hpsi}) and (\ref{ansatz2}) (we have set $\lambda =-1$)):
\bea
&& \mathcal{H}_\psi = \sum_{{\bm k}} \psi_{{\bm k},+}^\dagger \left(- \mu + \Theta^z  - 2 \sum_\rho w_\rho \Bigl[
\cos(\theta_\rho)  \cos( {\bm k} \cdot {\bm v}_\rho ) - \ell_\rho^z  \sin(\theta_\rho) \sin( {\bm k} \cdot {\bm v}_\rho  ) \Bigr] \right)  \psi_{{\bm k},+}^{\vphantom \dagger} \nn
&& ~~~~+ \sum_{{\bm k}} \psi_{{\bm k}+{\bm Q},-}^\dagger \left(- \mu - \Theta^z - 2 \sum_\rho w_\rho \Bigl[
\cos(\theta_\rho)  \cos( ({\bm k}+{\bm Q}) \cdot {\bm v}_\rho ) + \ell_\rho^z  \sin(\theta_\rho) \sin( ({\bm k}+{\bm Q}) \cdot {\bm v}_\rho  ) \Bigr] \right)  \psi_{{\bm k}+{\bm Q},-}^{\vphantom \dagger} \nn
&&+ \sum_{{\bm k}}  \psi_{{\bm k},+}^\dagger \psi_{{\bm k} + {\bm Q},-}^{\vphantom \dagger}\left( \Theta^x - i \Theta^y + \sum_\rho w_\rho \sin (\theta_\rho) (\ell_\rho^x - i \ell_\rho^y)
\Bigl[- i e^{i ({\bm k} + {\bm Q}) \cdot {\bm v}_\rho} + i e^{-i{\bm k}\cdot {\bm v}_\rho} \Bigr] \right)
+ \mbox{H.c.}.
\eea
The average kinetic energy and current on each bond can be evaluated from
\bea 
-\left\langle w_\rho  \psi_{i}^\dagger \, U^\rho (i) \, \psi_{i + {\bm v}_\rho}^{\vphantom \dagger} \right\rangle &=& 
-w_\rho \Bigl[
\cos(\theta_\rho)  + i \ell_\rho^z  \sin(\theta_\rho) \Bigr] \sum_{{\bm k}} e^{i {\bm k} \cdot {\bm v}_\rho}  
 \left\langle \psi_{{\bm k},+}^\dagger \psi_{{\bm k},+}^{\vphantom \dagger} \right\rangle \nn
 &~& -w_\rho \Bigl[
\cos(\theta_\rho)  - i \ell_\rho^z  \sin(\theta_\rho) \Bigr] \sum_{{\bm k}} e^{i ({\bm k} + {\bm Q}) \cdot {\bm v}_\rho}  
 \left\langle \psi_{{\bm k}+{\bm Q},-}^\dagger \psi_{{\bm k}+{\bm Q},-}^{\vphantom \dagger} \right\rangle \nn
&~& - i w_\rho \sin (\theta_\rho) (\ell_\rho^x - i \ell_\rho^y) \sum_{{\bm k}} e^{i ({\bm k} + {\bm Q}) \cdot {\bm v}_\rho}
\left\langle \psi_{{\bm k},+}^\dagger  \psi_{{\bm k} + {\bm Q},-}^{\vphantom \dagger} \right\rangle \nn
&~& - i w_\rho \sin (\theta_\rho) (\ell_\rho^x + i \ell_\rho^y) \sum_{{\bm k}} e^{i {\bm k} \cdot {\bm v}_\rho}
\left\langle \psi_{{\bm k}+ {\bm Q},-}^\dagger  \psi_{{\bm k},+ }^{\vphantom \dagger} \right\rangle.
\eea
Note that the result is explicitly independent of the site $i$.
The kinetic energy is twice the real part of the result, while the current, ${\bm J}_\rho$ in Eq.~(\ref{Jrho}), is $-{\bm v}_\rho$ times the imaginary part.

From the expression in momentum space, it is straightforward to see that the value of $\langle {\bf J}_\rho \rangle$ always obeys Bloch's theorem. The Hamiltonian $H_{\psi}$ can be re-written in momentum space in terms of a 2-component spinor $\chi_{\k}$ as follows:
\bea
H_{\psi} = \sum_{\k} \chi_{\k}^\dagger \, h_{\k} \, \chi_{\k}, \text{ where }
\chi_{\k} = \begin{pmatrix}
\psi_{\k,+} \\
\psi_{\k + \Q, - }
\end{pmatrix}.
\eea
The minimal coupling to an external electromagnetic gauge field $\bm{A}$ corresponds to a transformation $\k \rightarrow \k - \bm{A}$ in the momentum space Hamiltonian $h_{\k}$. The operator for the net current ${\bf J}$ in any given direction is the negative of the derivative with respect to $\bm{A}$ at $\bm A = 0$, which can be recast as a derivative with respect to $\k$:
\bea
\sum_{\rho} \langle {\bf J}_\rho \rangle = 
\sum_{\rho} \int \frac{d^2 k}{(2\pi)^2}\, \langle {\bf J}_{\rho}(\k) \rangle = \int \frac{d^2 k}{(2\pi)^2} \langle \chi_{\k}^\dagger \frac{\partial h_{\k}}{\partial \k} \chi_{\k}^{\vphantom\dagger}\rangle = \int \frac{d^2 k}{(2\pi)^2} \frac{\partial}{\partial \k} \langle \chi_{\k}^\dagger h_{\k} \chi_{\k}^{\vphantom\dagger}\rangle = 0,
\eea
where the last step uses the Feynman-Hellman theorem and periodicity in the Brillouin zone.

\section{Relation between loop currents and Higgs condensate}
\label{app:Higgs}

We show that a non-zero current necessarily requires a Higgs condensate. To do so, we need an operator which reverses the current, and is a symmetry of the Hamiltonian only if the Higgs condensate is absent. Consider the following anti-unitary operator $T$ that leaves the Higgs field unchanged:
\bea
T \, \psi_{s,\k} \, T^{-1} = (- i \tau^y)_{s, s^{\prime}} \, \psi_{s^{\prime},-\k}, ~~ T \, i  \,T^{-1} = -i, ~~ T \, H_i  \,T^{-1} = H_i.
\eea
Note that $T$ is not equivalent to the physical time-reversal $\mathcal{T}$ defined earlier in Eq.~(\ref{timesu2}), which always leaves $H_{\psi}$ invariant. Rather, as we show below, $T$ leaves the Hamiltonian invariant only if the Higgs condensate is absent.

Under $T$, we find the following transformation of the Hamiltonian $H_{\psi}$:
\bea
 T H_{\psi}T^{-1} = H_{\psi} - \sum_{\k} \chi^{\dagger}_{\k,s} \hat{\Theta}_{s,s^{\prime}} \, \chi_{\k,s^\prime}, \text{ where } \hat{\Theta} = 2 \begin{pmatrix}
\Theta^z & \Theta^- \\
\Theta^+ & -\Theta^z 
\end{pmatrix} = 2 \, \Theta^b \tau^b.
\eea
Therefore, the Hamiltonian $H_{\psi}$ commutes with $T$ when $\hat{\Theta} =0$. One can also show that the charge current operator $J_{\rho}(i)$ is odd under $T$, i.e,
\bea
T \, J_{\rho}(i) \, T^{-1} = - J_{\rho}(i).
\eea
Therefore, when $\hat{\Theta} = 0$, we can use the symmetry of $H_{\psi}$ under $T$ to find that:
\bea
\langle J_{\rho}(i) \rangle  = \langle T \, J_{\rho}(i) T^{-1} \rangle = - \langle J_{\rho}(i) \rangle \implies \langle J_{\rho}(i) \rangle = 0.
\eea
The physical content of the above equation is that current loop order cannot arise if all the SU(2) gauge bosons are deconfined, but can possibly arise when a Higgs condensate reduces the gauge group to U(1) or $\mathbb{Z}_2$.

\section{Real space perturbation theory for current in presence of large Higgs field}
\label{app:realspace}

We consider the limit where the Higgs field $H^a(i)$ is much larger compared to the hopping matrix elements of the $\psi_\pm$ fermions, characterized by $w_{\rho} U_{i, i + \vr}$. In the $|H_i| \rightarrow \infty$ limit, the Hamiltonian has only on-site terms, and therefore there is no current on the links. In this section, we perform a perturbation series expansion in $1/|H_i|$ to find an expression for the current. Recall that the charge gap of the SU(2) lattice gauge theory
is determined by $|H_i|$, and so this is similar to a $1/U$ expansion in the underlying Hubbard model.

We define a lattice Green's function in imaginary time in the standard fashion:
 \beq
 G_{ij}(\tau) = - \langle T_{\tau} (\psi_{i}(\tau) \psi_j^\dagger(0)) \rangle.
 \eeq
The Matsubara Green's function in the bare limit, $G^0_{i,n}$ is diagonal in real space (we set $\lambda = -1$):
\beq
G^0_{i,n} = (i\on - H_i^a \sa)^{-1} = \frac{i\on + H_i^a \sa}{(i \on)^2 - H_i^2}.
\eeq
 where we have set $\mu = 0$ for convenience (it does not modify our conclusion).
The Dyson equation for the Green's function in real space is given by:
 \bea
 G_{ij,n} &=& G^0_{i,n} \, \delta_{ij} + \sum_{k} G^0_{i,n} w_{ki}  U_{ki} G_{kj,n} \nn
 &=& G^0_{i,n} \, \delta_{ij} +  G^0_{i,n}  w_{ji} U_{ji} G^0_{j,n} + \sum_{k} G^0_{i,n} w_{ki} U_{ki} G^0_{k,n} w_{jk} U_{jk} G^{0}_{j,n} + ... \nn
 &\equiv& G^0_{i,n} \, \delta_{ij} + G^{(1)}_{ij,n} + G^{(2)}_{ij,n} + ...
 \eea
 where $w_{ij} = w_\rho$ is the hopping along the link $\langle i,j \rangle = \langle i, i + \vr \rangle$. Recall that the current operator  on the link $\langle i, i + \vr \rangle$ is given by:
\beq
J_{i, i + \vr} =  - i \vr w_\rho \left(  \psi_{i}^\dagger  \, U_{i,i+\vr} \, \psi_{i + {\bm v}_\rho}^{\vphantom \dagger}  - \mbox{H.c.} \right).
\eeq
Therefore, we can write the its expectation value in terms of the Green's function defined above as follows:
\beq
\langle J_{i, i + \vr}  \rangle  = - i \vr w_\rho \left[ \Tr(U_{i,i+\vr} G_{i + \vr,i}) - \Tr(U_{i+\vr,i} G_{i,i + \vr}) \right](\tau \rightarrow 0^-).
\eeq
The lowest order term in $1/|H_i|$ corresponds to $G_{ij} = G^0_{i}\delta_{ij}$, which gives zero current consistent with our expectations. To the next order in $1/|H_i|$, we find that the forward and backward currents exactly cancel and therefore the current is equal to zero to this order as well (via the cyclic property of the trace). 
\bea
\langle J^{(1)}_{i, i + \vr}  \rangle &=& - i \vr w_\rho \left[ \Tr(U_{i,i+\vr} G^{(1)}_{i + \vr,i}) - \Tr(U_{i+\vr,i} G^{(1)}_{i,i + \vr}) \right] \nn 
 &=& - i \vr w_\rho  \left[ \Tr(U_{i,i+\vr} G^0_{i + \vr} U_{i + \vr} G^0_i) - \Tr(U_{i+\vr,i} G^0_i U_{i,i + \vr} G^0_{i + \vr}) \right] = 0.
\eea
This exemplifies the importance of requiring non-nearest neighbor coupling for a non-zero current on the nearest neighbor bonds, albeit in a large Higgs field limit. 

The next term in the perturbation series, coming from $G^{(2)}$, gives us a non-zero current. To be more specific, let us label the sites as in Fig.~\ref{fig:O}, and compute the current from $m$ to $j$. It involves all triangles consisting of $m$, $j$ and a third site connected to both by a non-zero hopping (for simplicity we consider only nearest neighbor and next nearest neighbor hoppings ).
\bea
\label{eq:J}
\langle \bm{J}^{(2)}_{j,m}  \rangle &=& i \bm{v}_{jm} w_{1}^2 w_2 \Tr \left[ U_{jm} G^0_{m} \left( U_{ml} G^0_l U_{lj}  +  U_{mi} G^0_i U_{ij}  + U_{mn} G^0_n U_{nj} + U_{mk} G^0_k U_{kj} \right) G^0_j \right] \nn
& & -  i \bm{v}_{jm} w_{1}^2 w_2 \Tr \left[ U_{mj} G^0_j \left( U_{jl} G^0_l U_{lm} + U_{ji} G^0_i U_{im} + U_{jn} G^0_n U_{nm}+ U^{jk} G^0_k U_{km} \right ). G^0_{m}\right]
\eea
where $w_1, w_2$ are the nearest and next nearest neighbor hopping respectively. Note that the second term is just the Hermitian conjugate of the first term. We now convert to Matsubara Green's functions and evaluate the frequency summation (for simplicity we assume that $H_i$ are different on each site). The eigenstates at site $i$ have energy $\pm |H_i|$, therefore in the $T = 0$ limit only the negative energy eigenstates contribute to the current. 

Since the contribution of all triangular plaquettes to the current are similar, we only evaluate the contributions to the current by the first term in Eq.~(\ref{eq:J}) and its Hermitian conjugate (corresponding to the triangular plaquette $\triangle_{jlm}$). 
\bea
&& \frac{1}{\beta}\left( \sum_{i \on}  \Tr \left[ U_{jm} G^0_{m} U_{ml} G^0_l U_{lj} G^0_{j} \right] - \Tr \left[U_{jm} G^0_{m} U_{ml} G^0_l U_{lj}  G^0_j \right] \right) = \nn  
&&~~~~~\Tr \left[ U_{jm} \left( \frac{-|H_m| + H^a_m \sa}{-2 |H_m|} \right) U_{ml} \left( \frac{-|H_m| + H^a_l \sa}{H_m^2 - H_l^2} \right) U_{lj}  \left( \frac{-|H_m| + H^a_j \sa}{H_m^2 - H_j^2} \right) \right] \nn
&&~~~~~~~~~ + (j \rightarrow m \rightarrow l \rightarrow j) + (j \rightarrow l \rightarrow m \rightarrow j) - \mbox{H.c.}
\eea
Using the unitarity of U and $U_{\alpha \beta} = U^{\dagger}_{\beta \alpha}$, we can show quite generally that $\Tr(U_{jm} U_{ml} U_{lj}) = \Tr(U_{jl} U_{lm} U_{mj})$, so the term without any Higgs field $H_{\alpha}^a \sigma^a$ for some site $\alpha$ cancels with the contribution from the second line in Eq.~(\ref{eq:J}). The terms with two Higgs fields of the form $H_{\alpha}^a \sigma^a$ also cancel out with their Hermitian conjugates for the same reason. Therefore, we are left with two kinds of terms, both of which fall off as $|H_i|^{-2}$. The contribution to the current from this particular triangular plaquette can be written as:
\bea
J_{\triangle_{jlm}}  \sim & \; i \biggl[ \left( H^a_m \Tr(U_{jm}  \sa U_{ml} U_{lj}) + H^a_l  \Tr(U_{jm} U_{ml} \sa U_{lj})  + H^a_j  \Tr(U_{jm} U_{ml} U_{lj} \sa) \right) f(|H|) + \nn
 & H_m^a H_l^b H_j^c\Tr(U_{jm}\sa U_{ml} \sigma^b U_{lj} \sigma^c) g(|H|) -  \mbox{H.c.} \biggr],
\eea
where $f(|H|)$ and $g(|H|)$ are scalar functions of the Higgs fields (invariant under all symmetry operations), given by:
\bea
 f(|H|) &=&  \frac{1}{2(|H_m| + |H_l| )(|H_m| + |H_j|)(|H_m| + |H_l|)} \nn
 g(|H|) &=& -\frac{|H_m| + |H_{l}| + |H_{j}|}{2 |H_m| |H_l| |H_j| (|H_m| + |H_l| )(|H_m| + |H_j|)(|H_m| + |H_l|)}. 
\eea

We chose a particular set of terms, coming from $\triangle_{jlm}$ and the three other triangles related to it by reflection symmetries, as our order parameter $O_{jm}$ in Fig.~\ref{fig:O}. The other three contributions to the current on the link $\langle j,m \rangle$ in Eq.~(\ref{eq:J}) may be obtained by replacing the third vertex $l$ of the triangle by that of the triangle under consideration ($i,k$ and $n$). The net result for the current therefore contains the expressions presented in Fig.~\ref{fig:O}, along with three other expressions which can also serve as valid representations of ${\bm O}$.

\bibliography{pseudogap}

\begin{thebibliography}{65}%
\makeatletter
\providecommand \@ifxundefined [1]{%
 \@ifx{#1\undefined}
}%
\providecommand \@ifnum [1]{%
 \ifnum #1\expandafter \@firstoftwo
 \else \expandafter \@secondoftwo
 \fi
}%
\providecommand \@ifx [1]{%
 \ifx #1\expandafter \@firstoftwo
 \else \expandafter \@secondoftwo
 \fi
}%
\providecommand \natexlab [1]{#1}%
\providecommand \enquote  [1]{``#1''}%
\providecommand \bibnamefont  [1]{#1}%
\providecommand \bibfnamefont [1]{#1}%
\providecommand \citenamefont [1]{#1}%
\providecommand \href@noop [0]{\@secondoftwo}%
\providecommand \href [0]{\begingroup \@sanitize@url \@href}%
\providecommand \@href[1]{\@@startlink{#1}\@@href}%
\providecommand \@@href[1]{\endgroup#1\@@endlink}%
\providecommand \@sanitize@url [0]{\catcode `\\12\catcode `\$12\catcode
  `\&12\catcode `\#12\catcode `\^12\catcode `\_12\catcode `\%12\relax}%
\providecommand \@@startlink[1]{}%
\providecommand \@@endlink[0]{}%
\providecommand \url  [0]{\begingroup\@sanitize@url \@url }%
\providecommand \@url [1]{\endgroup\@href {#1}{\urlprefix }}%
\providecommand \urlprefix  [0]{URL }%
\providecommand \Eprint [0]{\href }%
\providecommand \doibase [0]{http://dx.doi.org/}%
\providecommand \selectlanguage [0]{\@gobble}%
\providecommand \bibinfo  [0]{\@secondoftwo}%
\providecommand \bibfield  [0]{\@secondoftwo}%
\providecommand \translation [1]{[#1]}%
\providecommand \BibitemOpen [0]{}%
\providecommand \bibitemStop [0]{}%
\providecommand \bibitemNoStop [0]{.\EOS\space}%
\providecommand \EOS [0]{\spacefactor3000\relax}%
\providecommand \BibitemShut  [1]{\csname bibitem#1\endcsname}%
\let\auto@bib@innerbib\@empty
\bibitem [{\citenamefont {{Ando}}\ \emph {et~al.}(2002)\citenamefont {{Ando}},
  \citenamefont {{Segawa}}, \citenamefont {{Komiya}},\ and\ \citenamefont
  {{Lavrov}}}]{2002PhRvL..88m7005A}%
  \BibitemOpen
  \bibfield  {author} {\bibinfo {author} {\bibfnamefont {Y.}~\bibnamefont
  {{Ando}}}, \bibinfo {author} {\bibfnamefont {K.}~\bibnamefont {{Segawa}}},
  \bibinfo {author} {\bibfnamefont {S.}~\bibnamefont {{Komiya}}}, \ and\
  \bibinfo {author} {\bibfnamefont {A.~N.}\ \bibnamefont {{Lavrov}}},\
  }\bibfield  {title} {\enquote {\bibinfo {title} {{Electrical Resistivity
  Anisotropy from Self-Organized One Dimensionality in High-Temperature
  Superconductors}},}\ }\href {\doibase 10.1103/PhysRevLett.88.137005}
  {\bibfield  {journal} {\bibinfo  {journal} {Phys. Rev. Lett.}\ }\textbf
  {\bibinfo {volume} {88}},\ \bibinfo {eid} {137005} (\bibinfo {year}
  {2002})},\ \Eprint {http://arxiv.org/abs/cond-mat/0108053} {cond-mat/0108053}
  \BibitemShut {NoStop}%
\bibitem [{\citenamefont {{Fauqu{\'e}}}\ \emph {et~al.}(2006)\citenamefont
  {{Fauqu{\'e}}}, \citenamefont {{Sidis}}, \citenamefont {{Hinkov}},
  \citenamefont {{Pailh{\`e}s}}, \citenamefont {{Lin}}, \citenamefont
  {{Chaud}},\ and\ \citenamefont {{Bourges}}}]{2006PhRvL..96s7001F}%
  \BibitemOpen
  \bibfield  {author} {\bibinfo {author} {\bibfnamefont {B.}~\bibnamefont
  {{Fauqu{\'e}}}}, \bibinfo {author} {\bibfnamefont {Y.}~\bibnamefont
  {{Sidis}}}, \bibinfo {author} {\bibfnamefont {V.}~\bibnamefont {{Hinkov}}},
  \bibinfo {author} {\bibfnamefont {S.}~\bibnamefont {{Pailh{\`e}s}}}, \bibinfo
  {author} {\bibfnamefont {C.~T.}\ \bibnamefont {{Lin}}}, \bibinfo {author}
  {\bibfnamefont {X.}~\bibnamefont {{Chaud}}}, \ and\ \bibinfo {author}
  {\bibfnamefont {P.}~\bibnamefont {{Bourges}}},\ }\bibfield  {title} {\enquote
  {\bibinfo {title} {{Magnetic Order in the Pseudogap Phase of High-T$_{c}$
  Superconductors}},}\ }\href {\doibase 10.1103/PhysRevLett.96.197001}
  {\bibfield  {journal} {\bibinfo  {journal} {Phys. Rev. Lett.}\ }\textbf
  {\bibinfo {volume} {96}},\ \bibinfo {eid} {197001} (\bibinfo {year}
  {2006})},\ \Eprint {http://arxiv.org/abs/cond-mat/0509210} {cond-mat/0509210}
  \BibitemShut {NoStop}%
\bibitem [{\citenamefont {Hinkov}\ \emph {et~al.}(2008)\citenamefont {Hinkov},
  \citenamefont {Haug}, \citenamefont {Fauqu{\'e}}, \citenamefont {Bourges},
  \citenamefont {Sidis}, \citenamefont {Ivanov}, \citenamefont {Bernhard},
  \citenamefont {Lin},\ and\ \citenamefont {Keimer}}]{Hinkov597}%
  \BibitemOpen
  \bibfield  {author} {\bibinfo {author} {\bibfnamefont {V.}~\bibnamefont
  {Hinkov}}, \bibinfo {author} {\bibfnamefont {D.}~\bibnamefont {Haug}},
  \bibinfo {author} {\bibfnamefont {B.}~\bibnamefont {Fauqu{\'e}}}, \bibinfo
  {author} {\bibfnamefont {P.}~\bibnamefont {Bourges}}, \bibinfo {author}
  {\bibfnamefont {Y.}~\bibnamefont {Sidis}}, \bibinfo {author} {\bibfnamefont
  {A.}~\bibnamefont {Ivanov}}, \bibinfo {author} {\bibfnamefont
  {C.}~\bibnamefont {Bernhard}}, \bibinfo {author} {\bibfnamefont {C.~T.}\
  \bibnamefont {Lin}}, \ and\ \bibinfo {author} {\bibfnamefont
  {B.}~\bibnamefont {Keimer}},\ }\bibfield  {title} {\enquote {\bibinfo {title}
  {{Electronic Liquid Crystal State in the High-Temperature Superconductor
  YBa$_2$Cu$_3$O$_{6.45}$}},}\ }\href {\doibase 10.1126/science.1152309}
  {\bibfield  {journal} {\bibinfo  {journal} {Science}\ }\textbf {\bibinfo
  {volume} {319}},\ \bibinfo {pages} {597} (\bibinfo {year}
  {2008})}\BibitemShut {NoStop}%
\bibitem [{\citenamefont {{Li}}\ \emph {et~al.}(2008)\citenamefont {{Li}},
  \citenamefont {{Bal{\'e}dent}}, \citenamefont {{Bari{\v s}i{\'c}}},
  \citenamefont {{Cho}}, \citenamefont {{Fauqu{\'e}}}, \citenamefont {{Sidis}},
  \citenamefont {{Yu}}, \citenamefont {{Zhao}}, \citenamefont {{Bourges}},\
  and\ \citenamefont {{Greven}}}]{2008arXiv0805.2959L}%
  \BibitemOpen
  \bibfield  {author} {\bibinfo {author} {\bibfnamefont {Y.}~\bibnamefont
  {{Li}}}, \bibinfo {author} {\bibfnamefont {V.}~\bibnamefont
  {{Bal{\'e}dent}}}, \bibinfo {author} {\bibfnamefont {N.}~\bibnamefont
  {{Bari{\v s}i{\'c}}}}, \bibinfo {author} {\bibfnamefont {Y.}~\bibnamefont
  {{Cho}}}, \bibinfo {author} {\bibfnamefont {B.}~\bibnamefont {{Fauqu{\'e}}}},
  \bibinfo {author} {\bibfnamefont {Y.}~\bibnamefont {{Sidis}}}, \bibinfo
  {author} {\bibfnamefont {G.}~\bibnamefont {{Yu}}}, \bibinfo {author}
  {\bibfnamefont {X.}~\bibnamefont {{Zhao}}}, \bibinfo {author} {\bibfnamefont
  {P.}~\bibnamefont {{Bourges}}}, \ and\ \bibinfo {author} {\bibfnamefont
  {M.}~\bibnamefont {{Greven}}},\ }\bibfield  {title} {\enquote {\bibinfo
  {title} {{Unusual magnetic order in the pseudogap region of the
  superconductor HgBa$_2$CuO$_{4+\delta}$}},}\ }\href {\doibase
  10.1038/nature07251} {\bibfield  {journal} {\bibinfo  {journal} {Nature}\
  }\textbf {\bibinfo {volume} {455}},\ \bibinfo {pages} {372} (\bibinfo {year}
  {2008})},\ \Eprint {http://arxiv.org/abs/0805.2959} {arXiv:0805.2959
  [cond-mat.supr-con]} \BibitemShut {NoStop}%
\bibitem [{\citenamefont {{Xia}}\ \emph {et~al.}(2008)\citenamefont {{Xia}},
  \citenamefont {{Schemm}}, \citenamefont {{Deutscher}}, \citenamefont
  {{Kivelson}}, \citenamefont {{Bonn}}, \citenamefont {{Hardy}}, \citenamefont
  {{Liang}}, \citenamefont {{Siemons}}, \citenamefont {{Koster}}, \citenamefont
  {{Fejer}},\ and\ \citenamefont {{Kapitulnik}}}]{2008PhRvL.100l7002X}%
  \BibitemOpen
  \bibfield  {author} {\bibinfo {author} {\bibfnamefont {J.}~\bibnamefont
  {{Xia}}}, \bibinfo {author} {\bibfnamefont {E.}~\bibnamefont {{Schemm}}},
  \bibinfo {author} {\bibfnamefont {G.}~\bibnamefont {{Deutscher}}}, \bibinfo
  {author} {\bibfnamefont {S.~A.}\ \bibnamefont {{Kivelson}}}, \bibinfo
  {author} {\bibfnamefont {D.~A.}\ \bibnamefont {{Bonn}}}, \bibinfo {author}
  {\bibfnamefont {W.~N.}\ \bibnamefont {{Hardy}}}, \bibinfo {author}
  {\bibfnamefont {R.}~\bibnamefont {{Liang}}}, \bibinfo {author} {\bibfnamefont
  {W.}~\bibnamefont {{Siemons}}}, \bibinfo {author} {\bibfnamefont
  {G.}~\bibnamefont {{Koster}}}, \bibinfo {author} {\bibfnamefont {M.~M.}\
  \bibnamefont {{Fejer}}}, \ and\ \bibinfo {author} {\bibfnamefont
  {A.}~\bibnamefont {{Kapitulnik}}},\ }\bibfield  {title} {\enquote {\bibinfo
  {title} {{Polar Kerr-Effect Measurements of the High-Temperature
  YBa$_{2}$Cu$_{3}$O$_{6+x}$ Superconductor: Evidence for Broken Symmetry near
  the Pseudogap Temperature}},}\ }\href {\doibase
  10.1103/PhysRevLett.100.127002} {\bibfield  {journal} {\bibinfo  {journal}
  {Phys. Rev. Lett.}\ }\textbf {\bibinfo {volume} {100}},\ \bibinfo {eid}
  {127002} (\bibinfo {year} {2008})},\ \Eprint {http://arxiv.org/abs/0711.2494}
  {arXiv:0711.2494 [cond-mat.supr-con]} \BibitemShut {NoStop}%
\bibitem [{\citenamefont {{Daou}}\ \emph {et~al.}(2010)\citenamefont {{Daou}},
  \citenamefont {{Chang}}, \citenamefont {{Leboeuf}}, \citenamefont
  {{Cyr-Choini{\`e}re}}, \citenamefont {{Lalibert{\'e}}}, \citenamefont
  {{Doiron-Leyraud}}, \citenamefont {{Ramshaw}}, \citenamefont {{Liang}},
  \citenamefont {{Bonn}}, \citenamefont {{Hardy}},\ and\ \citenamefont
  {{Taillefer}}}]{2010Natur.463..519D}%
  \BibitemOpen
  \bibfield  {author} {\bibinfo {author} {\bibfnamefont {R.}~\bibnamefont
  {{Daou}}}, \bibinfo {author} {\bibfnamefont {J.}~\bibnamefont {{Chang}}},
  \bibinfo {author} {\bibfnamefont {D.}~\bibnamefont {{Leboeuf}}}, \bibinfo
  {author} {\bibfnamefont {O.}~\bibnamefont {{Cyr-Choini{\`e}re}}}, \bibinfo
  {author} {\bibfnamefont {F.}~\bibnamefont {{Lalibert{\'e}}}}, \bibinfo
  {author} {\bibfnamefont {N.}~\bibnamefont {{Doiron-Leyraud}}}, \bibinfo
  {author} {\bibfnamefont {B.~J.}\ \bibnamefont {{Ramshaw}}}, \bibinfo {author}
  {\bibfnamefont {R.}~\bibnamefont {{Liang}}}, \bibinfo {author} {\bibfnamefont
  {D.~A.}\ \bibnamefont {{Bonn}}}, \bibinfo {author} {\bibfnamefont {W.~N.}\
  \bibnamefont {{Hardy}}}, \ and\ \bibinfo {author} {\bibfnamefont
  {L.}~\bibnamefont {{Taillefer}}},\ }\bibfield  {title} {\enquote {\bibinfo
  {title} {{Broken rotational symmetry in the pseudogap phase of a high-T$_{c}$
  superconductor}},}\ }\href {\doibase 10.1038/nature08716} {\bibfield
  {journal} {\bibinfo  {journal} {Nature}\ }\textbf {\bibinfo {volume} {463}},\
  \bibinfo {pages} {519} (\bibinfo {year} {2010})},\ \Eprint
  {http://arxiv.org/abs/0909.4430} {arXiv:0909.4430 [cond-mat.supr-con]}
  \BibitemShut {NoStop}%
\bibitem [{\citenamefont {{Li}}\ \emph {et~al.}(2010)\citenamefont {{Li}},
  \citenamefont {{Bal{\'e}dent}}, \citenamefont {{Yu}}, \citenamefont {{Bari{\v
  s}i{\'c}}}, \citenamefont {{Hradil}}, \citenamefont {{Mole}}, \citenamefont
  {{Sidis}}, \citenamefont {{Steffens}}, \citenamefont {{Zhao}}, \citenamefont
  {{Bourges}},\ and\ \citenamefont {{Greven}}}]{2010Natur.468..283L}%
  \BibitemOpen
  \bibfield  {author} {\bibinfo {author} {\bibfnamefont {Y.}~\bibnamefont
  {{Li}}}, \bibinfo {author} {\bibfnamefont {V.}~\bibnamefont
  {{Bal{\'e}dent}}}, \bibinfo {author} {\bibfnamefont {G.}~\bibnamefont
  {{Yu}}}, \bibinfo {author} {\bibfnamefont {N.}~\bibnamefont {{Bari{\v
  s}i{\'c}}}}, \bibinfo {author} {\bibfnamefont {K.}~\bibnamefont {{Hradil}}},
  \bibinfo {author} {\bibfnamefont {R.~A.}\ \bibnamefont {{Mole}}}, \bibinfo
  {author} {\bibfnamefont {Y.}~\bibnamefont {{Sidis}}}, \bibinfo {author}
  {\bibfnamefont {P.}~\bibnamefont {{Steffens}}}, \bibinfo {author}
  {\bibfnamefont {X.}~\bibnamefont {{Zhao}}}, \bibinfo {author} {\bibfnamefont
  {P.}~\bibnamefont {{Bourges}}}, \ and\ \bibinfo {author} {\bibfnamefont
  {M.}~\bibnamefont {{Greven}}},\ }\bibfield  {title} {\enquote {\bibinfo
  {title} {{Hidden magnetic excitation in the pseudogap phase of a high-T$_{c}$
  superconductor}},}\ }\href {\doibase 10.1038/nature09477} {\bibfield
  {journal} {\bibinfo  {journal} {Nature}\ }\textbf {\bibinfo {volume} {468}},\
  \bibinfo {pages} {283} (\bibinfo {year} {2010})},\ \Eprint
  {http://arxiv.org/abs/1007.2501} {arXiv:1007.2501 [cond-mat.supr-con]}
  \BibitemShut {NoStop}%
\bibitem [{\citenamefont {{Lawler}}\ \emph {et~al.}(2010)\citenamefont
  {{Lawler}}, \citenamefont {{Fujita}}, \citenamefont {{Lee}}, \citenamefont
  {{Schmidt}}, \citenamefont {{Kohsaka}}, \citenamefont {{Kim}}, \citenamefont
  {{Eisaki}}, \citenamefont {{Uchida}}, \citenamefont {{Davis}}, \citenamefont
  {{Sethna}},\ and\ \citenamefont {{Kim}}}]{2010Natur.466..347L}%
  \BibitemOpen
  \bibfield  {author} {\bibinfo {author} {\bibfnamefont {M.~J.}\ \bibnamefont
  {{Lawler}}}, \bibinfo {author} {\bibfnamefont {K.}~\bibnamefont {{Fujita}}},
  \bibinfo {author} {\bibfnamefont {J.}~\bibnamefont {{Lee}}}, \bibinfo
  {author} {\bibfnamefont {A.~R.}\ \bibnamefont {{Schmidt}}}, \bibinfo {author}
  {\bibfnamefont {Y.}~\bibnamefont {{Kohsaka}}}, \bibinfo {author}
  {\bibfnamefont {C.~K.}\ \bibnamefont {{Kim}}}, \bibinfo {author}
  {\bibfnamefont {H.}~\bibnamefont {{Eisaki}}}, \bibinfo {author}
  {\bibfnamefont {S.}~\bibnamefont {{Uchida}}}, \bibinfo {author}
  {\bibfnamefont {J.~C.}\ \bibnamefont {{Davis}}}, \bibinfo {author}
  {\bibfnamefont {J.~P.}\ \bibnamefont {{Sethna}}}, \ and\ \bibinfo {author}
  {\bibfnamefont {E.-A.}\ \bibnamefont {{Kim}}},\ }\bibfield  {title} {\enquote
  {\bibinfo {title} {{Intra-unit-cell electronic nematicity of the high-T$_{c}$
  copper-oxide pseudogap states}},}\ }\href {\doibase 10.1038/nature09169}
  {\bibfield  {journal} {\bibinfo  {journal} {Nature}\ }\textbf {\bibinfo
  {volume} {466}},\ \bibinfo {pages} {347} (\bibinfo {year} {2010})},\ \Eprint
  {http://arxiv.org/abs/1007.3216} {arXiv:1007.3216 [cond-mat.supr-con]}
  \BibitemShut {NoStop}%
\bibitem [{\citenamefont {{Lubashevsky}}\ \emph {et~al.}(2014)\citenamefont
  {{Lubashevsky}}, \citenamefont {{Pan}}, \citenamefont {{Kirzhner}},
  \citenamefont {{Koren}},\ and\ \citenamefont
  {{Armitage}}}]{2014PhRvL.112n7001L}%
  \BibitemOpen
  \bibfield  {author} {\bibinfo {author} {\bibfnamefont {Y.}~\bibnamefont
  {{Lubashevsky}}}, \bibinfo {author} {\bibfnamefont {L.}~\bibnamefont
  {{Pan}}}, \bibinfo {author} {\bibfnamefont {T.}~\bibnamefont {{Kirzhner}}},
  \bibinfo {author} {\bibfnamefont {G.}~\bibnamefont {{Koren}}}, \ and\
  \bibinfo {author} {\bibfnamefont {N.~P.}\ \bibnamefont {{Armitage}}},\
  }\bibfield  {title} {\enquote {\bibinfo {title} {{Optical Birefringence and
  Dichroism of Cuprate Superconductors in the THz Regime}},}\ }\href {\doibase
  10.1103/PhysRevLett.112.147001} {\bibfield  {journal} {\bibinfo  {journal}
  {Phys. Rev. Lett.}\ }\textbf {\bibinfo {volume} {112}},\ \bibinfo {eid}
  {147001} (\bibinfo {year} {2014})},\ \Eprint {http://arxiv.org/abs/1310.2265}
  {arXiv:1310.2265 [cond-mat.str-el]} \BibitemShut {NoStop}%
\bibitem [{\citenamefont {{Mangin-Thro}}\ \emph {et~al.}(2015)\citenamefont
  {{Mangin-Thro}}, \citenamefont {{Sidis}}, \citenamefont {{Wildes}},\ and\
  \citenamefont {{Bourges}}}]{2015NatCo...6E7705M}%
  \BibitemOpen
  \bibfield  {author} {\bibinfo {author} {\bibfnamefont {L.}~\bibnamefont
  {{Mangin-Thro}}}, \bibinfo {author} {\bibfnamefont {Y.}~\bibnamefont
  {{Sidis}}}, \bibinfo {author} {\bibfnamefont {A.}~\bibnamefont {{Wildes}}}, \
  and\ \bibinfo {author} {\bibfnamefont {P.}~\bibnamefont {{Bourges}}},\
  }\bibfield  {title} {\enquote {\bibinfo {title} {{Intra-unit-cell magnetic
  correlations near optimal doping in YBa$_{2}$Cu$_{3}$O$_{6.85}$}},}\ }\href
  {\doibase 10.1038/ncomms8705} {\bibfield  {journal} {\bibinfo  {journal}
  {Nature Communications}\ }\textbf {\bibinfo {volume} {6}},\ \bibinfo {eid}
  {7705} (\bibinfo {year} {2015})},\ \Eprint {http://arxiv.org/abs/1501.04919}
  {arXiv:1501.04919 [cond-mat.supr-con]} \BibitemShut {NoStop}%
\bibitem [{\citenamefont {{Zhao}}\ \emph {et~al.}(2016)\citenamefont {{Zhao}},
  \citenamefont {{Torchinsky}}, \citenamefont {{Chu}}, \citenamefont
  {{Ivanov}}, \citenamefont {{Lifshitz}}, \citenamefont {{Flint}},
  \citenamefont {{Qi}}, \citenamefont {{Cao}},\ and\ \citenamefont
  {{Hsieh}}}]{2016NatPh..12...32Z}%
  \BibitemOpen
  \bibfield  {author} {\bibinfo {author} {\bibfnamefont {L.}~\bibnamefont
  {{Zhao}}}, \bibinfo {author} {\bibfnamefont {D.~H.}\ \bibnamefont
  {{Torchinsky}}}, \bibinfo {author} {\bibfnamefont {H.}~\bibnamefont {{Chu}}},
  \bibinfo {author} {\bibfnamefont {V.}~\bibnamefont {{Ivanov}}}, \bibinfo
  {author} {\bibfnamefont {R.}~\bibnamefont {{Lifshitz}}}, \bibinfo {author}
  {\bibfnamefont {R.}~\bibnamefont {{Flint}}}, \bibinfo {author} {\bibfnamefont
  {T.}~\bibnamefont {{Qi}}}, \bibinfo {author} {\bibfnamefont {G.}~\bibnamefont
  {{Cao}}}, \ and\ \bibinfo {author} {\bibfnamefont {D.}~\bibnamefont
  {{Hsieh}}},\ }\bibfield  {title} {\enquote {\bibinfo {title} {{Evidence of an
  odd-parity hidden order in a spin-orbit coupled correlated iridate}},}\
  }\href {\doibase 10.1038/nphys3517} {\bibfield  {journal} {\bibinfo
  {journal} {Nature Physics}\ }\textbf {\bibinfo {volume} {12}},\ \bibinfo
  {pages} {32} (\bibinfo {year} {2016})},\ \Eprint
  {http://arxiv.org/abs/1601.01688} {arXiv:1601.01688 [cond-mat.str-el]}
  \BibitemShut {NoStop}%
\bibitem [{\citenamefont {{Zhao}}\ \emph {et~al.}(2017)\citenamefont {{Zhao}},
  \citenamefont {{Belvin}}, \citenamefont {{Liang}}, \citenamefont {{Bonn}},
  \citenamefont {{Hardy}}, \citenamefont {{Armitage}},\ and\ \citenamefont
  {{Hsieh}}}]{2016arXiv161108603Z}%
  \BibitemOpen
  \bibfield  {author} {\bibinfo {author} {\bibfnamefont {L.}~\bibnamefont
  {{Zhao}}}, \bibinfo {author} {\bibfnamefont {C.~A.}\ \bibnamefont
  {{Belvin}}}, \bibinfo {author} {\bibfnamefont {R.}~\bibnamefont {{Liang}}},
  \bibinfo {author} {\bibfnamefont {D.~A.}\ \bibnamefont {{Bonn}}}, \bibinfo
  {author} {\bibfnamefont {W.~N.}\ \bibnamefont {{Hardy}}}, \bibinfo {author}
  {\bibfnamefont {N.~P.}\ \bibnamefont {{Armitage}}}, \ and\ \bibinfo {author}
  {\bibfnamefont {D.}~\bibnamefont {{Hsieh}}},\ }\bibfield  {title} {\enquote
  {\bibinfo {title} {{A global inversion-symmetry-broken phase inside the
  pseudogap region of YBa$_2$Cu$_3$O$_y$}},}\ }\href
  {http://dx.doi.org/10.1038/nphys3962} {\bibfield  {journal} {\bibinfo
  {journal} {Nature Physics}\ }\textbf {\bibinfo {volume} {13}},\ \bibinfo
  {pages} {250} (\bibinfo {year} {2017})},\ \Eprint
  {http://arxiv.org/abs/1611.08603} {arXiv:1611.08603 [cond-mat.str-el]}
  \BibitemShut {NoStop}%
\bibitem [{\citenamefont {{Jeong}}\ \emph {et~al.}(2017)\citenamefont
  {{Jeong}}, \citenamefont {{Sidis}}, \citenamefont {{Louat}}, \citenamefont
  {{Brouet}},\ and\ \citenamefont {{Bourges}}}]{2017arXiv170106485J}%
  \BibitemOpen
  \bibfield  {author} {\bibinfo {author} {\bibfnamefont {J.}~\bibnamefont
  {{Jeong}}}, \bibinfo {author} {\bibfnamefont {Y.}~\bibnamefont {{Sidis}}},
  \bibinfo {author} {\bibfnamefont {A.}~\bibnamefont {{Louat}}}, \bibinfo
  {author} {\bibfnamefont {V.}~\bibnamefont {{Brouet}}}, \ and\ \bibinfo
  {author} {\bibfnamefont {P.}~\bibnamefont {{Bourges}}},\ }\bibfield  {title}
  {\enquote {\bibinfo {title} {{Time-reversal symmetry breaking hidden order in
  Sr$_2$(Ir,Rh)O$_4$}},}\ }\href {\doibase 10.1038/ncomms15119} {\bibfield
  {journal} {\bibinfo  {journal} {Nature Communications}\ }\textbf {\bibinfo
  {volume} {8}},\ \bibinfo {pages} {15119} (\bibinfo {year} {2017})},\ \Eprint
  {http://arxiv.org/abs/1701.06485} {arXiv:1701.06485 [cond-mat.str-el]}
  \BibitemShut {NoStop}%
\bibitem [{\citenamefont {Kivelson}\ \emph {et~al.}(1998)\citenamefont
  {Kivelson}, \citenamefont {Fradkin},\ and\ \citenamefont {Emery}}]{kfe98}%
  \BibitemOpen
  \bibfield  {author} {\bibinfo {author} {\bibfnamefont {S.~A.}\ \bibnamefont
  {Kivelson}}, \bibinfo {author} {\bibfnamefont {E.}~\bibnamefont {Fradkin}}, \
  and\ \bibinfo {author} {\bibfnamefont {V.~J.}\ \bibnamefont {Emery}},\
  }\bibfield  {title} {\enquote {\bibinfo {title} {{Electronic liquid-crystal
  phases of a doped Mott insulator}},}\ }\href@noop {} {\bibfield  {journal}
  {\bibinfo  {journal} {Nature}\ }\textbf {\bibinfo {volume} {393}},\ \bibinfo
  {pages} {550} (\bibinfo {year} {1998})}\BibitemShut {NoStop}%
\bibitem [{\citenamefont {{Varma}}(1997)}]{1997PhRvB..5514554V}%
  \BibitemOpen
  \bibfield  {author} {\bibinfo {author} {\bibfnamefont {C.~M.}\ \bibnamefont
  {{Varma}}},\ }\bibfield  {title} {\enquote {\bibinfo {title}
  {{Non-Fermi-liquid states and pairing instability of a general model of
  copper oxide metals}},}\ }\href {\doibase 10.1103/PhysRevB.55.14554}
  {\bibfield  {journal} {\bibinfo  {journal} {Phys. Rev. B}\ }\textbf {\bibinfo
  {volume} {55}},\ \bibinfo {pages} {14554} (\bibinfo {year} {1997})},\ \Eprint
  {http://arxiv.org/abs/cond-mat/9607105} {cond-mat/9607105} \BibitemShut
  {NoStop}%
\bibitem [{\citenamefont {{Simon}}\ and\ \citenamefont
  {{Varma}}(2002)}]{2002PhRvL..89x7003S}%
  \BibitemOpen
  \bibfield  {author} {\bibinfo {author} {\bibfnamefont {M.~E.}\ \bibnamefont
  {{Simon}}}\ and\ \bibinfo {author} {\bibfnamefont {C.~M.}\ \bibnamefont
  {{Varma}}},\ }\bibfield  {title} {\enquote {\bibinfo {title} {{Detection and
  Implications of a Time-Reversal Breaking State in Underdoped Cuprates}},}\
  }\href {\doibase 10.1103/PhysRevLett.89.247003} {\bibfield  {journal}
  {\bibinfo  {journal} {Phys. Rev. Lett.}\ }\textbf {\bibinfo {volume} {89}},\
  \bibinfo {eid} {247003} (\bibinfo {year} {2002})},\ \Eprint
  {http://arxiv.org/abs/cond-mat/0201036} {cond-mat/0201036} \BibitemShut
  {NoStop}%
\bibitem [{\citenamefont {{Simon}}\ and\ \citenamefont
  {{Varma}}(2003)}]{2003PhRvB..67e4511S}%
  \BibitemOpen
  \bibfield  {author} {\bibinfo {author} {\bibfnamefont {M.~E.}\ \bibnamefont
  {{Simon}}}\ and\ \bibinfo {author} {\bibfnamefont {C.~M.}\ \bibnamefont
  {{Varma}}},\ }\bibfield  {title} {\enquote {\bibinfo {title} {{Symmetry
  considerations for the detection of second-harmonic generation in cuprates in
  the pseudogap phase}},}\ }\href {\doibase 10.1103/PhysRevB.67.054511}
  {\bibfield  {journal} {\bibinfo  {journal} {Phys. Rev. B}\ }\textbf {\bibinfo
  {volume} {67}},\ \bibinfo {eid} {054511} (\bibinfo {year} {2003})},\ \Eprint
  {http://arxiv.org/abs/cond-mat/0210672} {cond-mat/0210672} \BibitemShut
  {NoStop}%
\bibitem [{\citenamefont {{Wang}}\ and\ \citenamefont
  {{Senthil}}(2011)}]{2011PhRvL.106m6402W}%
  \BibitemOpen
  \bibfield  {author} {\bibinfo {author} {\bibfnamefont {F.}~\bibnamefont
  {{Wang}}}\ and\ \bibinfo {author} {\bibfnamefont {T.}~\bibnamefont
  {{Senthil}}},\ }\bibfield  {title} {\enquote {\bibinfo {title} {{Twisted
  Hubbard Model for Sr$_{2}$IrO$_{4}$: Magnetism and Possible High Temperature
  Superconductivity}},}\ }\href {\doibase 10.1103/PhysRevLett.106.136402}
  {\bibfield  {journal} {\bibinfo  {journal} {Phys. Rev. Lett.}\ }\textbf
  {\bibinfo {volume} {106}},\ \bibinfo {eid} {136402} (\bibinfo {year}
  {2011})},\ \Eprint {http://arxiv.org/abs/1011.3500} {arXiv:1011.3500
  [cond-mat.str-el]} \BibitemShut {NoStop}%
\bibitem [{\citenamefont {Hertz}(1976)}]{hertz}%
  \BibitemOpen
  \bibfield  {author} {\bibinfo {author} {\bibfnamefont {J.~A.}\ \bibnamefont
  {Hertz}},\ }\bibfield  {title} {\enquote {\bibinfo {title} {{Quantum critical
  phenomena}},}\ }\href {\doibase 10.1103/PhysRevB.14.1165} {\bibfield
  {journal} {\bibinfo  {journal} {Phys. Rev. B}\ }\textbf {\bibinfo {volume}
  {14}},\ \bibinfo {pages} {1165} (\bibinfo {year} {1976})}\BibitemShut
  {NoStop}%
\bibitem [{\citenamefont {Chubukov}\ and\ \citenamefont
  {Sachdev}(1993)}]{CScomment}%
  \BibitemOpen
  \bibfield  {author} {\bibinfo {author} {\bibfnamefont {A.~V.}\ \bibnamefont
  {Chubukov}}\ and\ \bibinfo {author} {\bibfnamefont {S.}~\bibnamefont
  {Sachdev}},\ }\bibfield  {title} {\enquote {\bibinfo {title} {{Chubukov and
  Sachdev reply}},}\ }\href {\doibase 10.1103/PhysRevLett.71.3615} {\bibfield
  {journal} {\bibinfo  {journal} {Phys. Rev. Lett.}\ }\textbf {\bibinfo
  {volume} {71}},\ \bibinfo {pages} {3615} (\bibinfo {year}
  {1993})}\BibitemShut {NoStop}%
\bibitem [{\citenamefont {{Senthil}}\ \emph {et~al.}(2003)\citenamefont
  {{Senthil}}, \citenamefont {{Sachdev}},\ and\ \citenamefont {{Vojta}}}]{FFL}%
  \BibitemOpen
  \bibfield  {author} {\bibinfo {author} {\bibfnamefont {T.}~\bibnamefont
  {{Senthil}}}, \bibinfo {author} {\bibfnamefont {S.}~\bibnamefont
  {{Sachdev}}}, \ and\ \bibinfo {author} {\bibfnamefont {M.}~\bibnamefont
  {{Vojta}}},\ }\bibfield  {title} {\enquote {\bibinfo {title} {{Fractionalized
  Fermi Liquids}},}\ }\href {\doibase 10.1103/PhysRevLett.90.216403} {\bibfield
   {journal} {\bibinfo  {journal} {Phys. Rev. Lett.}\ }\textbf {\bibinfo
  {volume} {90}},\ \bibinfo {eid} {216403} (\bibinfo {year} {2003})},\ \Eprint
  {http://arxiv.org/abs/cond-mat/0209144} {cond-mat/0209144} \BibitemShut
  {NoStop}%
\bibitem [{\citenamefont {{Senthil}}\ \emph
  {et~al.}(2004{\natexlab{a}})\citenamefont {{Senthil}}, \citenamefont
  {{Vojta}},\ and\ \citenamefont {{Sachdev}}}]{TSMVSS04}%
  \BibitemOpen
  \bibfield  {author} {\bibinfo {author} {\bibfnamefont {T.}~\bibnamefont
  {{Senthil}}}, \bibinfo {author} {\bibfnamefont {M.}~\bibnamefont {{Vojta}}},
  \ and\ \bibinfo {author} {\bibfnamefont {S.}~\bibnamefont {{Sachdev}}},\
  }\bibfield  {title} {\enquote {\bibinfo {title} {{Weak magnetism and
  non-Fermi liquids near heavy-fermion critical points}},}\ }\href {\doibase
  10.1103/PhysRevB.69.035111} {\bibfield  {journal} {\bibinfo  {journal} {Phys.
  Rev. B}\ }\textbf {\bibinfo {volume} {69}},\ \bibinfo {eid} {035111}
  (\bibinfo {year} {2004}{\natexlab{a}})},\ \Eprint
  {http://arxiv.org/abs/cond-mat/0305193} {cond-mat/0305193} \BibitemShut
  {NoStop}%
\bibitem [{\citenamefont {{Paramekanti}}\ and\ \citenamefont
  {{Vishwanath}}(2004)}]{APAV04}%
  \BibitemOpen
  \bibfield  {author} {\bibinfo {author} {\bibfnamefont {A.}~\bibnamefont
  {{Paramekanti}}}\ and\ \bibinfo {author} {\bibfnamefont {A.}~\bibnamefont
  {{Vishwanath}}},\ }\bibfield  {title} {\enquote {\bibinfo {title} {{Extending
  Luttinger's theorem to $\mathbb{Z}_{2}$ fractionalized phases of matter}},}\
  }\href {\doibase 10.1103/PhysRevB.70.245118} {\bibfield  {journal} {\bibinfo
  {journal} {Phys. Rev. B}\ }\textbf {\bibinfo {volume} {70}},\ \bibinfo {eid}
  {245118} (\bibinfo {year} {2004})},\ \Eprint
  {http://arxiv.org/abs/cond-mat/0406619} {cond-mat/0406619} \BibitemShut
  {NoStop}%
\bibitem [{\citenamefont {{Punk}}\ \emph {et~al.}(2015)\citenamefont {{Punk}},
  \citenamefont {{Allais}},\ and\ \citenamefont
  {{Sachdev}}}]{2015PNAS..112.9552P}%
  \BibitemOpen
  \bibfield  {author} {\bibinfo {author} {\bibfnamefont {M.}~\bibnamefont
  {{Punk}}}, \bibinfo {author} {\bibfnamefont {A.}~\bibnamefont {{Allais}}}, \
  and\ \bibinfo {author} {\bibfnamefont {S.}~\bibnamefont {{Sachdev}}},\
  }\bibfield  {title} {\enquote {\bibinfo {title} {{Quantum dimer model for the
  pseudogap metal}},}\ }\href {\doibase 10.1073/pnas.1512206112} {\bibfield
  {journal} {\bibinfo  {journal} {Proc Nat. Acad. Sci}\ }\textbf {\bibinfo
  {volume} {112}},\ \bibinfo {pages} {9552} (\bibinfo {year} {2015})},\ \Eprint
  {http://arxiv.org/abs/1501.00978} {arXiv:1501.00978 [cond-mat.str-el]}
  \BibitemShut {NoStop}%
\bibitem [{\citenamefont {{Punk}}\ and\ \citenamefont
  {{Sachdev}}(2012)}]{2012PhRvB..85s5123P}%
  \BibitemOpen
  \bibfield  {author} {\bibinfo {author} {\bibfnamefont {M.}~\bibnamefont
  {{Punk}}}\ and\ \bibinfo {author} {\bibfnamefont {S.}~\bibnamefont
  {{Sachdev}}},\ }\bibfield  {title} {\enquote {\bibinfo {title} {{Fermi
  surface reconstruction in hole-doped $t$-$J$ models without long-range
  antiferromagnetic order}},}\ }\href {\doibase 10.1103/PhysRevB.85.195123}
  {\bibfield  {journal} {\bibinfo  {journal} {Phys. Rev. B}\ }\textbf {\bibinfo
  {volume} {85}},\ \bibinfo {eid} {195123} (\bibinfo {year} {2012})},\ \Eprint
  {http://arxiv.org/abs/1202.4023} {arXiv:1202.4023 [cond-mat.str-el]}
  \BibitemShut {NoStop}%
\bibitem [{Note1()}]{Note1}%
  \BibitemOpen
  \bibinfo {note} {Topological order is defined by the presence of ground state
  degeneracy of a system on a torus. More precisely, on a torus of size $L$,
  the lowest energy states have an energy difference which is of order
  $\protect \qopname \relax o{exp}(-\alpha L)$ for some constant $\alpha $.
  Topological order can also be present in gapless states, including those with
  Fermi surfaces \cite {FFL,TSMVSS04}. In such states, the non-topological
  gapless excitations have an energy of order $1/L^z$ (for some positive $z$)
  above the ground state on the torus, and so can be distinguished from the
  topologically degenerate states. Topological order is required for metals to
  have a Fermi surface volume distinct from the Luttinger volume \cite
  {TSMVSS04}, and hence to have a `pseudogap'.}\BibitemShut {Stop}%
\bibitem [{\citenamefont {{Sachdev}}\ and\ \citenamefont
  {{Chowdhury}}(2016)}]{DCSS16}%
  \BibitemOpen
  \bibfield  {author} {\bibinfo {author} {\bibfnamefont {S.}~\bibnamefont
  {{Sachdev}}}\ and\ \bibinfo {author} {\bibfnamefont {D.}~\bibnamefont
  {{Chowdhury}}},\ }\bibfield  {title} {\enquote {\bibinfo {title} {{The novel
  metallic states of the cuprates: Fermi liquids with topological order and
  strange metals}},}\ }\href {\doibase 10.1093/ptep/ptw110} {\bibfield
  {journal} {\bibinfo  {journal} {Prog. Theor. Exp. Phys.}\ }\textbf {\bibinfo
  {volume} {2016}},\ \bibinfo {eid} {12C102} (\bibinfo {year} {2016})},\
  \Eprint {http://arxiv.org/abs/1605.03579} {arXiv:1605.03579
  [cond-mat.str-el]} \BibitemShut {NoStop}%
\bibitem [{\citenamefont {Read}\ and\ \citenamefont {Sachdev}(1991)}]{NRSS91}%
  \BibitemOpen
  \bibfield  {author} {\bibinfo {author} {\bibfnamefont {N.}~\bibnamefont
  {Read}}\ and\ \bibinfo {author} {\bibfnamefont {S.}~\bibnamefont {Sachdev}},\
  }\bibfield  {title} {\enquote {\bibinfo {title} {{Large $N$ expansion for
  frustrated quantum antiferromagnets}},}\ }\href {\doibase
  10.1103/PhysRevLett.66.1773} {\bibfield  {journal} {\bibinfo  {journal}
  {Phys. Rev. Lett.}\ }\textbf {\bibinfo {volume} {66}},\ \bibinfo {pages}
  {1773} (\bibinfo {year} {1991})}\BibitemShut {NoStop}%
\bibitem [{\citenamefont {Sachdev}\ and\ \citenamefont {Read}(1991)}]{SSNR91}%
  \BibitemOpen
  \bibfield  {author} {\bibinfo {author} {\bibfnamefont {S.}~\bibnamefont
  {Sachdev}}\ and\ \bibinfo {author} {\bibfnamefont {N.}~\bibnamefont {Read}},\
  }\bibfield  {title} {\enquote {\bibinfo {title} {{Large $N$ expansion for
  frustrated and doped quantum antiferromagnets}},}\ }\href {\doibase
  10.1142/S0217979291000158} {\bibfield  {journal} {\bibinfo  {journal} {Int.
  J. Mod. Phys. B}\ }\textbf {\bibinfo {volume} {5}},\ \bibinfo {pages} {219}
  (\bibinfo {year} {1991})},\ \Eprint {http://arxiv.org/abs/cond-mat/0402109}
  {cond-mat/0402109} \BibitemShut {NoStop}%
\bibitem [{\citenamefont {{Barkeshli}}\ \emph {et~al.}(2013)\citenamefont
  {{Barkeshli}}, \citenamefont {{Yao}},\ and\ \citenamefont
  {{Kivelson}}}]{2013PhRvB..87n0402B}%
  \BibitemOpen
  \bibfield  {author} {\bibinfo {author} {\bibfnamefont {M.}~\bibnamefont
  {{Barkeshli}}}, \bibinfo {author} {\bibfnamefont {H.}~\bibnamefont {{Yao}}},
  \ and\ \bibinfo {author} {\bibfnamefont {S.~A.}\ \bibnamefont {{Kivelson}}},\
  }\bibfield  {title} {\enquote {\bibinfo {title} {{Gapless spin liquids:
  Stability and possible experimental relevance}},}\ }\href {\doibase
  10.1103/PhysRevB.87.140402} {\bibfield  {journal} {\bibinfo  {journal} {Phys.
  Rev. B}\ }\textbf {\bibinfo {volume} {87}},\ \bibinfo {eid} {140402}
  (\bibinfo {year} {2013})},\ \Eprint {http://arxiv.org/abs/1208.3869}
  {arXiv:1208.3869 [cond-mat.str-el]} \BibitemShut {NoStop}%
\bibitem [{\citenamefont {{Sachdev}}\ \emph {et~al.}(2009)\citenamefont
  {{Sachdev}}, \citenamefont {{Metlitski}}, \citenamefont {{Qi}},\ and\
  \citenamefont {{Xu}}}]{SS09}%
  \BibitemOpen
  \bibfield  {author} {\bibinfo {author} {\bibfnamefont {S.}~\bibnamefont
  {{Sachdev}}}, \bibinfo {author} {\bibfnamefont {M.~A.}\ \bibnamefont
  {{Metlitski}}}, \bibinfo {author} {\bibfnamefont {Y.}~\bibnamefont {{Qi}}}, \
  and\ \bibinfo {author} {\bibfnamefont {C.}~\bibnamefont {{Xu}}},\ }\bibfield
  {title} {\enquote {\bibinfo {title} {{Fluctuating spin density waves in
  metals}},}\ }\href {\doibase 10.1103/PhysRevB.80.155129} {\bibfield
  {journal} {\bibinfo  {journal} {Phys. Rev. B}\ }\textbf {\bibinfo {volume}
  {80}},\ \bibinfo {eid} {155129} (\bibinfo {year} {2009})},\ \Eprint
  {http://arxiv.org/abs/0907.3732} {arXiv:0907.3732 [cond-mat.str-el]}
  \BibitemShut {NoStop}%
\bibitem [{\citenamefont {{Chowdhury}}\ and\ \citenamefont
  {{Sachdev}}(2015)}]{DCSS15b}%
  \BibitemOpen
  \bibfield  {author} {\bibinfo {author} {\bibfnamefont {D.}~\bibnamefont
  {{Chowdhury}}}\ and\ \bibinfo {author} {\bibfnamefont {S.}~\bibnamefont
  {{Sachdev}}},\ }\bibfield  {title} {\enquote {\bibinfo {title} {{Higgs
  criticality in a two-dimensional metal}},}\ }\href {\doibase
  10.1103/PhysRevB.91.115123} {\bibfield  {journal} {\bibinfo  {journal} {Phys.
  Rev. B}\ }\textbf {\bibinfo {volume} {91}},\ \bibinfo {eid} {115123}
  (\bibinfo {year} {2015})},\ \Eprint {http://arxiv.org/abs/1412.1086}
  {arXiv:1412.1086 [cond-mat.str-el]} \BibitemShut {NoStop}%
\bibitem [{\citenamefont {Chakravarty}\ \emph {et~al.}(1988)\citenamefont
  {Chakravarty}, \citenamefont {Halperin},\ and\ \citenamefont
  {Nelson}}]{CHN1}%
  \BibitemOpen
  \bibfield  {author} {\bibinfo {author} {\bibfnamefont {S.}~\bibnamefont
  {Chakravarty}}, \bibinfo {author} {\bibfnamefont {B.~I.}\ \bibnamefont
  {Halperin}}, \ and\ \bibinfo {author} {\bibfnamefont {D.~R.}\ \bibnamefont
  {Nelson}},\ }\bibfield  {title} {\enquote {\bibinfo {title} {{Low-temperature
  behavior of two-dimensional quantum antiferromagnets}},}\ }\href {\doibase
  10.1103/PhysRevLett.60.1057} {\bibfield  {journal} {\bibinfo  {journal}
  {Phys. Rev. Lett.}\ }\textbf {\bibinfo {volume} {60}},\ \bibinfo {pages}
  {1057} (\bibinfo {year} {1988})}\BibitemShut {NoStop}%
\bibitem [{\citenamefont {Chakravarty}\ \emph {et~al.}(1989)\citenamefont
  {Chakravarty}, \citenamefont {Halperin},\ and\ \citenamefont
  {Nelson}}]{CHN2}%
  \BibitemOpen
  \bibfield  {author} {\bibinfo {author} {\bibfnamefont {S.}~\bibnamefont
  {Chakravarty}}, \bibinfo {author} {\bibfnamefont {B.~I.}\ \bibnamefont
  {Halperin}}, \ and\ \bibinfo {author} {\bibfnamefont {D.~R.}\ \bibnamefont
  {Nelson}},\ }\bibfield  {title} {\enquote {\bibinfo {title} {{Two-dimensional
  quantum Heisenberg antiferromagnet at low temperatures}},}\ }\href {\doibase
  10.1103/PhysRevB.39.2344} {\bibfield  {journal} {\bibinfo  {journal} {Phys.
  Rev. B}\ }\textbf {\bibinfo {volume} {39}},\ \bibinfo {pages} {2344}
  (\bibinfo {year} {1989})}\BibitemShut {NoStop}%
\bibitem [{\citenamefont {Haldane}(1988)}]{Haldane88}%
  \BibitemOpen
  \bibfield  {author} {\bibinfo {author} {\bibfnamefont {F.~D.~M.}\
  \bibnamefont {Haldane}},\ }\bibfield  {title} {\enquote {\bibinfo {title}
  {{O(3) Nonlinear $\sigma$ Model and the Topological Distinction between
  Integer- and Half-Integer-Spin Antiferromagnets in Two Dimensions}},}\ }\href
  {\doibase 10.1103/PhysRevLett.61.1029} {\bibfield  {journal} {\bibinfo
  {journal} {Phys. Rev. Lett.}\ }\textbf {\bibinfo {volume} {61}},\ \bibinfo
  {pages} {1029} (\bibinfo {year} {1988})}\BibitemShut {NoStop}%
\bibitem [{\citenamefont {Read}\ and\ \citenamefont {Sachdev}(1989)}]{NRSS89}%
  \BibitemOpen
  \bibfield  {author} {\bibinfo {author} {\bibfnamefont {N.}~\bibnamefont
  {Read}}\ and\ \bibinfo {author} {\bibfnamefont {S.}~\bibnamefont {Sachdev}},\
  }\bibfield  {title} {\enquote {\bibinfo {title} {{Valence-bond and
  spin-Peierls ground states of low-dimensional quantum antiferromagnets}},}\
  }\href {\doibase 10.1103/PhysRevLett.62.1694} {\bibfield  {journal} {\bibinfo
   {journal} {Phys. Rev. Lett.}\ }\textbf {\bibinfo {volume} {62}},\ \bibinfo
  {pages} {1694} (\bibinfo {year} {1989})}\BibitemShut {NoStop}%
\bibitem [{\citenamefont {Read}\ and\ \citenamefont {Sachdev}(1990)}]{NRSS90}%
  \BibitemOpen
  \bibfield  {author} {\bibinfo {author} {\bibfnamefont {N.}~\bibnamefont
  {Read}}\ and\ \bibinfo {author} {\bibfnamefont {S.}~\bibnamefont {Sachdev}},\
  }\bibfield  {title} {\enquote {\bibinfo {title} {{Spin-Peierls, valence-bond
  solid, and N\'eel ground states of low-dimensional quantum
  antiferromagnets}},}\ }\href {\doibase 10.1103/PhysRevB.42.4568} {\bibfield
  {journal} {\bibinfo  {journal} {Phys. Rev. B}\ }\textbf {\bibinfo {volume}
  {42}},\ \bibinfo {pages} {4568} (\bibinfo {year} {1990})}\BibitemShut
  {NoStop}%
\bibitem [{\citenamefont {{Senthil}}\ \emph
  {et~al.}(2004{\natexlab{b}})\citenamefont {{Senthil}}, \citenamefont
  {{Vishwanath}}, \citenamefont {{Balents}}, \citenamefont {{Sachdev}},\ and\
  \citenamefont {{Fisher}}}]{senthil1}%
  \BibitemOpen
  \bibfield  {author} {\bibinfo {author} {\bibfnamefont {T.}~\bibnamefont
  {{Senthil}}}, \bibinfo {author} {\bibfnamefont {A.}~\bibnamefont
  {{Vishwanath}}}, \bibinfo {author} {\bibfnamefont {L.}~\bibnamefont
  {{Balents}}}, \bibinfo {author} {\bibfnamefont {S.}~\bibnamefont
  {{Sachdev}}}, \ and\ \bibinfo {author} {\bibfnamefont {M.~P.~A.}\
  \bibnamefont {{Fisher}}},\ }\bibfield  {title} {\enquote {\bibinfo {title}
  {{Deconfined Quantum Critical Points}},}\ }\href {\doibase
  10.1126/science.1091806} {\bibfield  {journal} {\bibinfo  {journal}
  {Science}\ }\textbf {\bibinfo {volume} {303}},\ \bibinfo {pages} {1490}
  (\bibinfo {year} {2004}{\natexlab{b}})},\ \Eprint
  {http://arxiv.org/abs/cond-mat/0311326} {cond-mat/0311326} \BibitemShut
  {NoStop}%
\bibitem [{\citenamefont {{Senthil}}\ \emph
  {et~al.}(2004{\natexlab{c}})\citenamefont {{Senthil}}, \citenamefont
  {{Balents}}, \citenamefont {{Sachdev}}, \citenamefont {{Vishwanath}},\ and\
  \citenamefont {{Fisher}}}]{senthil2}%
  \BibitemOpen
  \bibfield  {author} {\bibinfo {author} {\bibfnamefont {T.}~\bibnamefont
  {{Senthil}}}, \bibinfo {author} {\bibfnamefont {L.}~\bibnamefont
  {{Balents}}}, \bibinfo {author} {\bibfnamefont {S.}~\bibnamefont
  {{Sachdev}}}, \bibinfo {author} {\bibfnamefont {A.}~\bibnamefont
  {{Vishwanath}}}, \ and\ \bibinfo {author} {\bibfnamefont {M.~P.~A.}\
  \bibnamefont {{Fisher}}},\ }\bibfield  {title} {\enquote {\bibinfo {title}
  {{Quantum criticality beyond the Landau-Ginzburg-Wilson paradigm}},}\ }\href
  {\doibase 10.1103/PhysRevB.70.144407} {\bibfield  {journal} {\bibinfo
  {journal} {\prb}\ }\textbf {\bibinfo {volume} {70}},\ \bibinfo {eid} {144407}
  (\bibinfo {year} {2004}{\natexlab{c}})},\ \Eprint
  {http://arxiv.org/abs/cond-mat/0312617} {cond-mat/0312617} \BibitemShut
  {NoStop}%
\bibitem [{\citenamefont {Bohm}(1949)}]{Bohm49}%
  \BibitemOpen
  \bibfield  {author} {\bibinfo {author} {\bibfnamefont {D.}~\bibnamefont
  {Bohm}},\ }\bibfield  {title} {\enquote {\bibinfo {title} {{Note on a Theorem
  of Bloch Concerning Possible Causes of Superconductivity}},}\ }\href
  {\doibase 10.1103/PhysRev.75.502} {\bibfield  {journal} {\bibinfo  {journal}
  {Phys. Rev.}\ }\textbf {\bibinfo {volume} {75}},\ \bibinfo {pages} {502}
  (\bibinfo {year} {1949})}\BibitemShut {NoStop}%
\bibitem [{\citenamefont {{Ohashi}}\ and\ \citenamefont
  {{Momoi}}(1996)}]{1996JPSJ...65.3254O}%
  \BibitemOpen
  \bibfield  {author} {\bibinfo {author} {\bibfnamefont {Y.}~\bibnamefont
  {{Ohashi}}}\ and\ \bibinfo {author} {\bibfnamefont {T.}~\bibnamefont
  {{Momoi}}},\ }\bibfield  {title} {\enquote {\bibinfo {title} {{On the Bloch
  Theorem Concerning Spontaneous Electric Current}},}\ }\href {\doibase
  10.1143/JPSJ.65.3254} {\bibfield  {journal} {\bibinfo  {journal} {J. Phys.
  Soc. Jpn.}\ }\textbf {\bibinfo {volume} {65}},\ \bibinfo {pages} {3254}
  (\bibinfo {year} {1996})},\ \Eprint {http://arxiv.org/abs/cond-mat/9606182}
  {cond-mat/9606182} \BibitemShut {NoStop}%
\bibitem [{\citenamefont {{Stanescu}}\ and\ \citenamefont
  {{Phillips}}(2004)}]{2004PhRvB..69x5104S}%
  \BibitemOpen
  \bibfield  {author} {\bibinfo {author} {\bibfnamefont {T.~D.}\ \bibnamefont
  {{Stanescu}}}\ and\ \bibinfo {author} {\bibfnamefont {P.}~\bibnamefont
  {{Phillips}}},\ }\bibfield  {title} {\enquote {\bibinfo {title}
  {{Nonperturbative approach to full Mott behavior}},}\ }\href {\doibase
  10.1103/PhysRevB.69.245104} {\bibfield  {journal} {\bibinfo  {journal} {Phys.
  Rev. B}\ }\textbf {\bibinfo {volume} {69}},\ \bibinfo {eid} {245104}
  (\bibinfo {year} {2004})},\ \Eprint {http://arxiv.org/abs/cond-mat/0301254}
  {cond-mat/0301254} \BibitemShut {NoStop}%
\bibitem [{\citenamefont {{Berg}}\ \emph {et~al.}(2008)\citenamefont {{Berg}},
  \citenamefont {{Chen}},\ and\ \citenamefont
  {{Kivelson}}}]{2008PhRvL.100b7003B}%
  \BibitemOpen
  \bibfield  {author} {\bibinfo {author} {\bibfnamefont {E.}~\bibnamefont
  {{Berg}}}, \bibinfo {author} {\bibfnamefont {C.-C.}\ \bibnamefont {{Chen}}},
  \ and\ \bibinfo {author} {\bibfnamefont {S.~A.}\ \bibnamefont {{Kivelson}}},\
  }\bibfield  {title} {\enquote {\bibinfo {title} {{Stability of Nodal
  Quasiparticles in Superconductors with Coexisting Orders}},}\ }\href
  {\doibase 10.1103/PhysRevLett.100.027003} {\bibfield  {journal} {\bibinfo
  {journal} {Phys. Rev. Lett.}\ }\textbf {\bibinfo {volume} {100}},\ \bibinfo
  {eid} {027003} (\bibinfo {year} {2008})},\ \Eprint
  {http://arxiv.org/abs/0710.0113} {arXiv:0710.0113 [cond-mat.supr-con]}
  \BibitemShut {NoStop}%
\bibitem [{\citenamefont {{Sachdev}}\ \emph {et~al.}(2016)\citenamefont
  {{Sachdev}}, \citenamefont {{Berg}}, \citenamefont {{Chatterjee}},\ and\
  \citenamefont {{Schattner}}}]{2016PhRvB..94k5147S}%
  \BibitemOpen
  \bibfield  {author} {\bibinfo {author} {\bibfnamefont {S.}~\bibnamefont
  {{Sachdev}}}, \bibinfo {author} {\bibfnamefont {E.}~\bibnamefont {{Berg}}},
  \bibinfo {author} {\bibfnamefont {S.}~\bibnamefont {{Chatterjee}}}, \ and\
  \bibinfo {author} {\bibfnamefont {Y.}~\bibnamefont {{Schattner}}},\
  }\bibfield  {title} {\enquote {\bibinfo {title} {{Spin density wave order,
  topological order, and Fermi surface reconstruction}},}\ }\href {\doibase
  10.1103/PhysRevB.94.115147} {\bibfield  {journal} {\bibinfo  {journal} {Phys.
  Rev. B}\ }\textbf {\bibinfo {volume} {94}},\ \bibinfo {eid} {115147}
  (\bibinfo {year} {2016})},\ \Eprint {http://arxiv.org/abs/1606.07813}
  {arXiv:1606.07813 [cond-mat.str-el]} \BibitemShut {NoStop}%
\bibitem [{\citenamefont {Yoshida}\ \emph {et~al.}(2012)\citenamefont
  {Yoshida}, \citenamefont {Schr\"oder}, \citenamefont {Ferriani},
  \citenamefont {Serrate}, \citenamefont {Kubetzka}, \citenamefont {von
  Bergmann}, \citenamefont {Heinze},\ and\ \citenamefont
  {Wiesendanger}}]{PhysRevLett.108.087205}%
  \BibitemOpen
  \bibfield  {author} {\bibinfo {author} {\bibfnamefont {Y.}~\bibnamefont
  {Yoshida}}, \bibinfo {author} {\bibfnamefont {S.}~\bibnamefont {Schr\"oder}},
  \bibinfo {author} {\bibfnamefont {P.}~\bibnamefont {Ferriani}}, \bibinfo
  {author} {\bibfnamefont {D.}~\bibnamefont {Serrate}}, \bibinfo {author}
  {\bibfnamefont {A.}~\bibnamefont {Kubetzka}}, \bibinfo {author}
  {\bibfnamefont {K.}~\bibnamefont {von Bergmann}}, \bibinfo {author}
  {\bibfnamefont {S.}~\bibnamefont {Heinze}}, \ and\ \bibinfo {author}
  {\bibfnamefont {R.}~\bibnamefont {Wiesendanger}},\ }\bibfield  {title}
  {\enquote {\bibinfo {title} {{Conical Spin-Spiral State in an Ultrathin Film
  Driven by Higher-Order Spin Interactions}},}\ }\href {\doibase
  10.1103/PhysRevLett.108.087205} {\bibfield  {journal} {\bibinfo  {journal}
  {Phys. Rev. Lett.}\ }\textbf {\bibinfo {volume} {108}},\ \bibinfo {pages}
  {087205} (\bibinfo {year} {2012})}\BibitemShut {NoStop}%
\bibitem [{\citenamefont {{Hermele}}\ \emph {et~al.}(2004)\citenamefont
  {{Hermele}}, \citenamefont {{Senthil}}, \citenamefont {{Fisher}},
  \citenamefont {{Lee}}, \citenamefont {{Nagaosa}},\ and\ \citenamefont
  {{Wen}}}]{2004PhRvB..70u4437H}%
  \BibitemOpen
  \bibfield  {author} {\bibinfo {author} {\bibfnamefont {M.}~\bibnamefont
  {{Hermele}}}, \bibinfo {author} {\bibfnamefont {T.}~\bibnamefont
  {{Senthil}}}, \bibinfo {author} {\bibfnamefont {M.~P.~A.}\ \bibnamefont
  {{Fisher}}}, \bibinfo {author} {\bibfnamefont {P.~A.}\ \bibnamefont {{Lee}}},
  \bibinfo {author} {\bibfnamefont {N.}~\bibnamefont {{Nagaosa}}}, \ and\
  \bibinfo {author} {\bibfnamefont {X.-G.}\ \bibnamefont {{Wen}}},\ }\bibfield
  {title} {\enquote {\bibinfo {title} {{Stability of U (1) spin liquids in two
  dimensions}},}\ }\href {\doibase 10.1103/PhysRevB.70.214437} {\bibfield
  {journal} {\bibinfo  {journal} {Phys. Rev. B}\ }\textbf {\bibinfo {volume}
  {70}},\ \bibinfo {eid} {214437} (\bibinfo {year} {2004})},\ \Eprint
  {http://arxiv.org/abs/cond-mat/0404751} {cond-mat/0404751} \BibitemShut
  {NoStop}%
\bibitem [{\citenamefont {{Kaul}}\ \emph
  {et~al.}(2008{\natexlab{a}})\citenamefont {{Kaul}}, \citenamefont {{Kim}},
  \citenamefont {{Sachdev}},\ and\ \citenamefont
  {{Senthil}}}]{2008NatPh...4...28K}%
  \BibitemOpen
  \bibfield  {author} {\bibinfo {author} {\bibfnamefont {R.~K.}\ \bibnamefont
  {{Kaul}}}, \bibinfo {author} {\bibfnamefont {Y.~B.}\ \bibnamefont {{Kim}}},
  \bibinfo {author} {\bibfnamefont {S.}~\bibnamefont {{Sachdev}}}, \ and\
  \bibinfo {author} {\bibfnamefont {T.}~\bibnamefont {{Senthil}}},\ }\bibfield
  {title} {\enquote {\bibinfo {title} {{Algebraic charge liquids}},}\ }\href
  {\doibase 10.1038/nphys790} {\bibfield  {journal} {\bibinfo  {journal}
  {Nature Physics}\ }\textbf {\bibinfo {volume} {4}},\ \bibinfo {pages} {28}
  (\bibinfo {year} {2008}{\natexlab{a}})},\ \Eprint
  {http://arxiv.org/abs/0706.2187} {arXiv:0706.2187 [cond-mat.str-el]}
  \BibitemShut {NoStop}%
\bibitem [{\citenamefont {Wen}(1991)}]{PhysRevB.44.2664}%
  \BibitemOpen
  \bibfield  {author} {\bibinfo {author} {\bibfnamefont {X.~G.}\ \bibnamefont
  {Wen}},\ }\bibfield  {title} {\enquote {\bibinfo {title} {Mean-field theory
  of spin-liquid states with finite energy gap and topological orders},}\
  }\href {\doibase 10.1103/PhysRevB.44.2664} {\bibfield  {journal} {\bibinfo
  {journal} {Phys. Rev. B}\ }\textbf {\bibinfo {volume} {44}},\ \bibinfo
  {pages} {2664} (\bibinfo {year} {1991})}\BibitemShut {NoStop}%
\bibitem [{\citenamefont {{Kaul}}\ \emph {et~al.}(2007)\citenamefont {{Kaul}},
  \citenamefont {{Kolezhuk}}, \citenamefont {{Levin}}, \citenamefont
  {{Sachdev}},\ and\ \citenamefont {{Senthil}}}]{2007PhRvB..75w5122K}%
  \BibitemOpen
  \bibfield  {author} {\bibinfo {author} {\bibfnamefont {R.~K.}\ \bibnamefont
  {{Kaul}}}, \bibinfo {author} {\bibfnamefont {A.}~\bibnamefont {{Kolezhuk}}},
  \bibinfo {author} {\bibfnamefont {M.}~\bibnamefont {{Levin}}}, \bibinfo
  {author} {\bibfnamefont {S.}~\bibnamefont {{Sachdev}}}, \ and\ \bibinfo
  {author} {\bibfnamefont {T.}~\bibnamefont {{Senthil}}},\ }\bibfield  {title}
  {\enquote {\bibinfo {title} {{Hole dynamics in an antiferromagnet across a
  deconfined quantum critical point}},}\ }\href {\doibase
  10.1103/PhysRevB.75.235122} {\bibfield  {journal} {\bibinfo  {journal} {Phys.
  Rev. B}\ }\textbf {\bibinfo {volume} {75}},\ \bibinfo {eid} {235122}
  (\bibinfo {year} {2007})},\ \Eprint {http://arxiv.org/abs/cond-mat/0702119}
  {cond-mat/0702119} \BibitemShut {NoStop}%
\bibitem [{\citenamefont {{Kaul}}\ \emph
  {et~al.}(2008{\natexlab{b}})\citenamefont {{Kaul}}, \citenamefont
  {{Metlitski}}, \citenamefont {{Sachdev}},\ and\ \citenamefont
  {{Xu}}}]{2008PhRvB..78d5110K}%
  \BibitemOpen
  \bibfield  {author} {\bibinfo {author} {\bibfnamefont {R.~K.}\ \bibnamefont
  {{Kaul}}}, \bibinfo {author} {\bibfnamefont {M.~A.}\ \bibnamefont
  {{Metlitski}}}, \bibinfo {author} {\bibfnamefont {S.}~\bibnamefont
  {{Sachdev}}}, \ and\ \bibinfo {author} {\bibfnamefont {C.}~\bibnamefont
  {{Xu}}},\ }\bibfield  {title} {\enquote {\bibinfo {title} {{Destruction of
  N{\'e}el order in the cuprates by electron doping}},}\ }\href {\doibase
  10.1103/PhysRevB.78.045110} {\bibfield  {journal} {\bibinfo  {journal} {Phys.
  Rev. B}\ }\textbf {\bibinfo {volume} {78}},\ \bibinfo {eid} {045110}
  (\bibinfo {year} {2008}{\natexlab{b}})},\ \Eprint
  {http://arxiv.org/abs/0804.1794} {arXiv:0804.1794 [cond-mat.str-el]}
  \BibitemShut {NoStop}%
\bibitem [{\citenamefont {{Fernandes}}\ and\ \citenamefont
  {{Chubukov}}(2017)}]{RMFAC17}%
  \BibitemOpen
  \bibfield  {author} {\bibinfo {author} {\bibfnamefont {R.~M.}\ \bibnamefont
  {{Fernandes}}}\ and\ \bibinfo {author} {\bibfnamefont {A.~V.}\ \bibnamefont
  {{Chubukov}}},\ }\bibfield  {title} {\enquote {\bibinfo {title} {{Low-energy
  microscopic models for iron-based superconductors: a review}},}\ }\href
  {\doibase 10.1088/1361-6633/80/1/014503} {\bibfield  {journal} {\bibinfo
  {journal} {Rep. Prog. Phys.}\ }\textbf {\bibinfo {volume} {80}},\ \bibinfo
  {eid} {014503} (\bibinfo {year} {2017})},\ \Eprint
  {http://arxiv.org/abs/1607.00865} {arXiv:1607.00865 [cond-mat.str-el]}
  \BibitemShut {NoStop}%
\bibitem [{\citenamefont {{Badoux}}\ \emph {et~al.}(2016)\citenamefont
  {{Badoux}}, \citenamefont {{Tabis}}, \citenamefont {{Lalibert{\'e}}},
  \citenamefont {{Grissonnanche}}, \citenamefont {{Vignolle}}, \citenamefont
  {{Vignolles}}, \citenamefont {{B{\'e}ard}}, \citenamefont {{Bonn}},
  \citenamefont {{Hardy}}, \citenamefont {{Liang}}, \citenamefont
  {{Doiron-Leyraud}}, \citenamefont {{Taillefer}},\ and\ \citenamefont
  {{Proust}}}]{LTCP15}%
  \BibitemOpen
  \bibfield  {author} {\bibinfo {author} {\bibfnamefont {S.}~\bibnamefont
  {{Badoux}}}, \bibinfo {author} {\bibfnamefont {W.}~\bibnamefont {{Tabis}}},
  \bibinfo {author} {\bibfnamefont {F.}~\bibnamefont {{Lalibert{\'e}}}},
  \bibinfo {author} {\bibfnamefont {G.}~\bibnamefont {{Grissonnanche}}},
  \bibinfo {author} {\bibfnamefont {B.}~\bibnamefont {{Vignolle}}}, \bibinfo
  {author} {\bibfnamefont {D.}~\bibnamefont {{Vignolles}}}, \bibinfo {author}
  {\bibfnamefont {J.}~\bibnamefont {{B{\'e}ard}}}, \bibinfo {author}
  {\bibfnamefont {D.~A.}\ \bibnamefont {{Bonn}}}, \bibinfo {author}
  {\bibfnamefont {W.~N.}\ \bibnamefont {{Hardy}}}, \bibinfo {author}
  {\bibfnamefont {R.}~\bibnamefont {{Liang}}}, \bibinfo {author} {\bibfnamefont
  {N.}~\bibnamefont {{Doiron-Leyraud}}}, \bibinfo {author} {\bibfnamefont
  {L.}~\bibnamefont {{Taillefer}}}, \ and\ \bibinfo {author} {\bibfnamefont
  {C.}~\bibnamefont {{Proust}}},\ }\bibfield  {title} {\enquote {\bibinfo
  {title} {{Change of carrier density at the pseudogap critical point of a
  cuprate superconductor}},}\ }\href {\doibase 10.1038/nature16983} {\bibfield
  {journal} {\bibinfo  {journal} {Nature}\ }\textbf {\bibinfo {volume} {531}},\
  \bibinfo {pages} {210} (\bibinfo {year} {2016})},\ \Eprint
  {http://arxiv.org/abs/1511.08162} {arXiv:1511.08162 [cond-mat.supr-con]}
  \BibitemShut {NoStop}%
\bibitem [{\citenamefont {Cooper}\ \emph {et~al.}(2009)\citenamefont {Cooper},
  \citenamefont {Wang}, \citenamefont {Vignolle}, \citenamefont {Lipscombe},
  \citenamefont {Hayden}, \citenamefont {Tanabe}, \citenamefont {Adachi},
  \citenamefont {Koike}, \citenamefont {Nohara}, \citenamefont {Takagi},
  \citenamefont {Proust},\ and\ \citenamefont {Hussey}}]{Cooper603}%
  \BibitemOpen
  \bibfield  {author} {\bibinfo {author} {\bibfnamefont {R.~A.}\ \bibnamefont
  {Cooper}}, \bibinfo {author} {\bibfnamefont {Y.}~\bibnamefont {Wang}},
  \bibinfo {author} {\bibfnamefont {B.}~\bibnamefont {Vignolle}}, \bibinfo
  {author} {\bibfnamefont {O.~J.}\ \bibnamefont {Lipscombe}}, \bibinfo {author}
  {\bibfnamefont {S.~M.}\ \bibnamefont {Hayden}}, \bibinfo {author}
  {\bibfnamefont {Y.}~\bibnamefont {Tanabe}}, \bibinfo {author} {\bibfnamefont
  {T.}~\bibnamefont {Adachi}}, \bibinfo {author} {\bibfnamefont
  {Y.}~\bibnamefont {Koike}}, \bibinfo {author} {\bibfnamefont
  {M.}~\bibnamefont {Nohara}}, \bibinfo {author} {\bibfnamefont
  {H.}~\bibnamefont {Takagi}}, \bibinfo {author} {\bibfnamefont
  {C.}~\bibnamefont {Proust}}, \ and\ \bibinfo {author} {\bibfnamefont {N.~E.}\
  \bibnamefont {Hussey}},\ }\bibfield  {title} {\enquote {\bibinfo {title}
  {{Anomalous Criticality in the Electrical Resistivity of
  La$_{2-x}$Sr$_x$CuO$_4$}},}\ }\href {\doibase 10.1126/science.1165015}
  {\bibfield  {journal} {\bibinfo  {journal} {Science}\ }\textbf {\bibinfo
  {volume} {323}},\ \bibinfo {pages} {603} (\bibinfo {year}
  {2009})}\BibitemShut {NoStop}%
\bibitem [{\citenamefont {Bo{\v z}ovi{\'c}}\ \emph {et~al.}(2016)\citenamefont
  {Bo{\v z}ovi{\'c}}, \citenamefont {He}, \citenamefont {Wu},\ and\
  \citenamefont {Bollinger}}]{Bozovic}%
  \BibitemOpen
  \bibfield  {author} {\bibinfo {author} {\bibfnamefont {I.}~\bibnamefont
  {Bo{\v z}ovi{\'c}}}, \bibinfo {author} {\bibfnamefont {X.}~\bibnamefont
  {He}}, \bibinfo {author} {\bibfnamefont {J.}~\bibnamefont {Wu}}, \ and\
  \bibinfo {author} {\bibfnamefont {A.~T.}\ \bibnamefont {Bollinger}},\
  }\bibfield  {title} {\enquote {\bibinfo {title} {{Dependence of the critical
  temperature in overdoped copper oxides on superfluid density}},}\ }\href
  {\doibase 10.1038/nature19061} {\bibfield  {journal} {\bibinfo  {journal}
  {Nature}\ }\textbf {\bibinfo {volume} {536}},\ \bibinfo {pages} {309}
  (\bibinfo {year} {2016})}\BibitemShut {NoStop}%
\bibitem [{\citenamefont {{Wen}}(2002)}]{Wen2002}%
  \BibitemOpen
  \bibfield  {author} {\bibinfo {author} {\bibfnamefont {X.-G.}\ \bibnamefont
  {{Wen}}},\ }\bibfield  {title} {\enquote {\bibinfo {title} {{Quantum orders
  and symmetric spin liquids}},}\ }\href {\doibase 10.1103/PhysRevB.65.165113}
  {\bibfield  {journal} {\bibinfo  {journal} {Phys. Rev. B}\ }\textbf {\bibinfo
  {volume} {65}},\ \bibinfo {eid} {165113} (\bibinfo {year} {2002})},\ \Eprint
  {http://arxiv.org/abs/cond-mat/0107071} {cond-mat/0107071} \BibitemShut
  {NoStop}%
\bibitem [{\citenamefont {{Chubukov}}\ \emph {et~al.}(1994)\citenamefont
  {{Chubukov}}, \citenamefont {{Senthil}},\ and\ \citenamefont
  {{Sachdev}}}]{1994PhRvL..72.2089C}%
  \BibitemOpen
  \bibfield  {author} {\bibinfo {author} {\bibfnamefont {A.~V.}\ \bibnamefont
  {{Chubukov}}}, \bibinfo {author} {\bibfnamefont {T.}~\bibnamefont
  {{Senthil}}}, \ and\ \bibinfo {author} {\bibfnamefont {S.}~\bibnamefont
  {{Sachdev}}},\ }\bibfield  {title} {\enquote {\bibinfo {title} {{Universal
  magnetic properties of frustrated quantum antiferromagnets in two
  dimensions}},}\ }\href {\doibase 10.1103/PhysRevLett.72.2089} {\bibfield
  {journal} {\bibinfo  {journal} {Phys. Rev. Lett.}\ }\textbf {\bibinfo
  {volume} {72}},\ \bibinfo {pages} {2089} (\bibinfo {year} {1994})},\ \Eprint
  {http://arxiv.org/abs/cond-mat/9311045} {cond-mat/9311045} \BibitemShut
  {NoStop}%
\bibitem [{\citenamefont {{Sachdev}}(2003)}]{2003RvMP...75..913S}%
  \BibitemOpen
  \bibfield  {author} {\bibinfo {author} {\bibfnamefont {S.}~\bibnamefont
  {{Sachdev}}},\ }\bibfield  {title} {\enquote {\bibinfo {title} {{Colloquium:
  Order and quantum phase transitions in the cuprate superconductors}},}\
  }\href {\doibase 10.1103/RevModPhys.75.913} {\bibfield  {journal} {\bibinfo
  {journal} {Rev. Mod. Phys.}\ }\textbf {\bibinfo {volume} {75}},\ \bibinfo
  {pages} {913} (\bibinfo {year} {2003})},\ \Eprint
  {http://arxiv.org/abs/cond-mat/0211005} {cond-mat/0211005} \BibitemShut
  {NoStop}%
\bibitem [{\citenamefont {{Senthil}}\ and\ \citenamefont
  {{Fisher}}(2000)}]{SenthilFisher}%
  \BibitemOpen
  \bibfield  {author} {\bibinfo {author} {\bibfnamefont {T.}~\bibnamefont
  {{Senthil}}}\ and\ \bibinfo {author} {\bibfnamefont {M.~P.~A.}\ \bibnamefont
  {{Fisher}}},\ }\bibfield  {title} {\enquote {\bibinfo {title}
  {{$\mathbb{Z}_{2}$ gauge theory of electron fractionalization in strongly
  correlated systems}},}\ }\href {\doibase 10.1103/PhysRevB.62.7850} {\bibfield
   {journal} {\bibinfo  {journal} {Phys. Rev. B}\ }\textbf {\bibinfo {volume}
  {62}},\ \bibinfo {pages} {7850} (\bibinfo {year} {2000})},\ \Eprint
  {http://arxiv.org/abs/cond-mat/9910224} {cond-mat/9910224} \BibitemShut
  {NoStop}%
\bibitem [{\citenamefont {{Balents}}\ and\ \citenamefont
  {{Sachdev}}(2007)}]{2007AnPhy.322.2635B}%
  \BibitemOpen
  \bibfield  {author} {\bibinfo {author} {\bibfnamefont {L.}~\bibnamefont
  {{Balents}}}\ and\ \bibinfo {author} {\bibfnamefont {S.}~\bibnamefont
  {{Sachdev}}},\ }\bibfield  {title} {\enquote {\bibinfo {title} {{Dual vortex
  theory of doped Mott insulators}},}\ }\href {\doibase
  10.1016/j.aop.2007.02.001} {\bibfield  {journal} {\bibinfo  {journal} {Annals
  of Physics}\ }\textbf {\bibinfo {volume} {322}},\ \bibinfo {pages} {2635}
  (\bibinfo {year} {2007})},\ \Eprint {http://arxiv.org/abs/cond-mat/0612220}
  {cond-mat/0612220} \BibitemShut {NoStop}%
\bibitem [{\citenamefont {{Chatterjee}}\ \emph {et~al.}(2016)\citenamefont
  {{Chatterjee}}, \citenamefont {{Qi}}, \citenamefont {{Sachdev}},\ and\
  \citenamefont {{Steinberg}}}]{2016PhRvB..94b4502C}%
  \BibitemOpen
  \bibfield  {author} {\bibinfo {author} {\bibfnamefont {S.}~\bibnamefont
  {{Chatterjee}}}, \bibinfo {author} {\bibfnamefont {Y.}~\bibnamefont {{Qi}}},
  \bibinfo {author} {\bibfnamefont {S.}~\bibnamefont {{Sachdev}}}, \ and\
  \bibinfo {author} {\bibfnamefont {J.}~\bibnamefont {{Steinberg}}},\
  }\bibfield  {title} {\enquote {\bibinfo {title} {{Superconductivity from a
  confinement transition out of a fractionalized Fermi liquid with Z$_{2}$
  topological and Ising-nematic orders}},}\ }\href {\doibase
  10.1103/PhysRevB.94.024502} {\bibfield  {journal} {\bibinfo  {journal} {Phys.
  Rev. B}\ }\textbf {\bibinfo {volume} {94}},\ \bibinfo {eid} {024502}
  (\bibinfo {year} {2016})},\ \Eprint {http://arxiv.org/abs/1603.03041}
  {arXiv:1603.03041 [cond-mat.str-el]} \BibitemShut {NoStop}%
\bibitem [{\citenamefont {{Eberlein}}\ \emph {et~al.}(2016)\citenamefont
  {{Eberlein}}, \citenamefont {{Metzner}}, \citenamefont {{Sachdev}},\ and\
  \citenamefont {{Yamase}}}]{2016PhRvL.117r7001E}%
  \BibitemOpen
  \bibfield  {author} {\bibinfo {author} {\bibfnamefont {A.}~\bibnamefont
  {{Eberlein}}}, \bibinfo {author} {\bibfnamefont {W.}~\bibnamefont
  {{Metzner}}}, \bibinfo {author} {\bibfnamefont {S.}~\bibnamefont
  {{Sachdev}}}, \ and\ \bibinfo {author} {\bibfnamefont {H.}~\bibnamefont
  {{Yamase}}},\ }\bibfield  {title} {\enquote {\bibinfo {title} {{Fermi Surface
  Reconstruction and Drop in the Hall Number due to Spiral Antiferromagnetism
  in High-T$_{c}$ Cuprates}},}\ }\href {\doibase
  10.1103/PhysRevLett.117.187001} {\bibfield  {journal} {\bibinfo  {journal}
  {Phys. Rev. Lett.}\ }\textbf {\bibinfo {volume} {117}},\ \bibinfo {eid}
  {187001} (\bibinfo {year} {2016})},\ \Eprint
  {http://arxiv.org/abs/1607.06087} {arXiv:1607.06087 [cond-mat.str-el]}
  \BibitemShut {NoStop}%
\bibitem [{\citenamefont {Chatterjee}\ \emph {et~al.}(2017)\citenamefont
  {Chatterjee}, \citenamefont {Sachdev},\ and\ \citenamefont
  {Eberlein}}]{CSE17}%
  \BibitemOpen
  \bibfield  {author} {\bibinfo {author} {\bibfnamefont {S.}~\bibnamefont
  {Chatterjee}}, \bibinfo {author} {\bibfnamefont {S.}~\bibnamefont {Sachdev}},
  \ and\ \bibinfo {author} {\bibfnamefont {A.}~\bibnamefont {Eberlein}},\
  }\bibfield  {title} {\enquote {\bibinfo {title} {{Thermal and electrical
  transport in metals and superconductors across antiferromagnetic and
  topological quantum transitions}},}\ }\href@noop {} {\  (\bibinfo {year}
  {2017})},\ \Eprint {http://arxiv.org/abs/1704.02329} {arXiv:1704.02329
  [cond-mat.str-el]} \BibitemShut {NoStop}%
\bibitem [{\citenamefont {{Laliberte}}\ \emph {et~al.}(2016)\citenamefont
  {{Laliberte}}, \citenamefont {{Tabis}}, \citenamefont {{Badoux}},
  \citenamefont {{Vignolle}}, \citenamefont {{Destraz}}, \citenamefont
  {{Momono}}, \citenamefont {{Kurosawa}}, \citenamefont {{Yamada}},
  \citenamefont {{Takagi}}, \citenamefont {{Doiron-Leyraud}}, \citenamefont
  {{Proust}},\ and\ \citenamefont {{Taillefer}}}]{Laliberte2016-arXiv}%
  \BibitemOpen
  \bibfield  {author} {\bibinfo {author} {\bibfnamefont {F.}~\bibnamefont
  {{Laliberte}}}, \bibinfo {author} {\bibfnamefont {W.}~\bibnamefont
  {{Tabis}}}, \bibinfo {author} {\bibfnamefont {S.}~\bibnamefont {{Badoux}}},
  \bibinfo {author} {\bibfnamefont {B.}~\bibnamefont {{Vignolle}}}, \bibinfo
  {author} {\bibfnamefont {D.}~\bibnamefont {{Destraz}}}, \bibinfo {author}
  {\bibfnamefont {N.}~\bibnamefont {{Momono}}}, \bibinfo {author}
  {\bibfnamefont {T.}~\bibnamefont {{Kurosawa}}}, \bibinfo {author}
  {\bibfnamefont {K.}~\bibnamefont {{Yamada}}}, \bibinfo {author}
  {\bibfnamefont {H.}~\bibnamefont {{Takagi}}}, \bibinfo {author}
  {\bibfnamefont {N.}~\bibnamefont {{Doiron-Leyraud}}}, \bibinfo {author}
  {\bibfnamefont {C.}~\bibnamefont {{Proust}}}, \ and\ \bibinfo {author}
  {\bibfnamefont {L.}~\bibnamefont {{Taillefer}}},\ }\bibfield  {title}
  {\enquote {\bibinfo {title} {{Origin of the metal-to-insulator crossover in
  cuprate superconductors}},}\ }\href@noop {} {\bibfield  {journal} {\bibinfo
  {journal} {ArXiv e-prints}\ } (\bibinfo {year} {2016})},\ \Eprint
  {http://arxiv.org/abs/1606.04491} {arXiv:1606.04491 [cond-mat.supr-con]}
  \BibitemShut {NoStop}%
\bibitem [{\citenamefont {{Collignon}}\ \emph {et~al.}(2016)\citenamefont
  {{Collignon}}, \citenamefont {{Badoux}}, \citenamefont {{Afshar}},
  \citenamefont {{Michon}}, \citenamefont {{Laliberte}}, \citenamefont
  {{Cyr-Choiniere}}, \citenamefont {{Zhou}}, \citenamefont {{Licciardello}},
  \citenamefont {{Wiedmann}}, \citenamefont {{Doiron-Leyraud}},\ and\
  \citenamefont {{Taillefer}}}]{Collignon2016-arXiv}%
  \BibitemOpen
  \bibfield  {author} {\bibinfo {author} {\bibfnamefont {C.}~\bibnamefont
  {{Collignon}}}, \bibinfo {author} {\bibfnamefont {S.}~\bibnamefont
  {{Badoux}}}, \bibinfo {author} {\bibfnamefont {S.~A.~A.}\ \bibnamefont
  {{Afshar}}}, \bibinfo {author} {\bibfnamefont {B.}~\bibnamefont {{Michon}}},
  \bibinfo {author} {\bibfnamefont {F.}~\bibnamefont {{Laliberte}}}, \bibinfo
  {author} {\bibfnamefont {O.}~\bibnamefont {{Cyr-Choiniere}}}, \bibinfo
  {author} {\bibfnamefont {J.-S.}\ \bibnamefont {{Zhou}}}, \bibinfo {author}
  {\bibfnamefont {S.}~\bibnamefont {{Licciardello}}}, \bibinfo {author}
  {\bibfnamefont {S.}~\bibnamefont {{Wiedmann}}}, \bibinfo {author}
  {\bibfnamefont {N.}~\bibnamefont {{Doiron-Leyraud}}}, \ and\ \bibinfo
  {author} {\bibfnamefont {L.}~\bibnamefont {{Taillefer}}},\ }\bibfield
  {title} {\enquote {\bibinfo {title} {Fermi-surface transformation across the
  pseudogap critical point of the cuprate superconductor
  $\mathrm{La}_{1.6-x}\mathrm{Nd}_{0.4}\mathrm{Sr}_{x}\mathrm{CuO}_4$},}\
  }\href@noop {} {\bibfield  {journal} {\bibinfo  {journal} {ArXiv e-prints}\ }
  (\bibinfo {year} {2016})},\ \Eprint {http://arxiv.org/abs/1607.05693}
  {arXiv:1607.05693 [cond-mat.supr-con]} \BibitemShut {NoStop}%
\bibitem [{\citenamefont {Michon}\ \emph {et~al.}(2017)\citenamefont {Michon},
  \citenamefont {Li}, \citenamefont {Bourgeois-Hope}, \citenamefont {Badoux},
  \citenamefont {Zhou}, \citenamefont {Doiron-Leyraud},\ and\ \citenamefont
  {Taillefer}}]{Michon2017}%
  \BibitemOpen
  \bibfield  {author} {\bibinfo {author} {\bibfnamefont {B.~M.}\ \bibnamefont
  {Michon}}, \bibinfo {author} {\bibfnamefont {S.}~\bibnamefont {Li}}, \bibinfo
  {author} {\bibfnamefont {P.}~\bibnamefont {Bourgeois-Hope}}, \bibinfo
  {author} {\bibfnamefont {S.}~\bibnamefont {Badoux}}, \bibinfo {author}
  {\bibfnamefont {J.-S.}\ \bibnamefont {Zhou}}, \bibinfo {author}
  {\bibfnamefont {N.}~\bibnamefont {Doiron-Leyraud}}, \ and\ \bibinfo {author}
  {\bibfnamefont {L.}~\bibnamefont {Taillefer}},\ }\bibfield  {title} {\enquote
  {\bibinfo {title} {{Pseudogap critical point inside the superconducting phase
  of a cuprate : a thermal conductivity study in Nd-LSCO}},}\ }\href@noop {}
  {\bibfield  {journal} {\bibinfo  {journal} {unpublished}\ } (\bibinfo {year}
  {2017})}\BibitemShut {NoStop}%
\end{thebibliography}%

\end{document}